\documentclass[aps, prc, twocolumn, floatfix, nofootinbib, superscriptaddress, longbibliography, amsmath, amssymb]{revtex4-1}

\usepackage{mathrsfs}
\usepackage{graphicx}
\usepackage{epsfig}
\usepackage{bm}
\usepackage{bbold}
\usepackage{color}
\usepackage{float}
\usepackage{enumerate}
\usepackage{dcolumn}
\usepackage{multirow} 
\usepackage{hyperref}
\usepackage{color}
\usepackage{braket}
\usepackage{dcolumn}
\usepackage{multirow}
\usepackage[titletoc]{appendix}
\usepackage{cleveref}
\usepackage{xfrac}
\usepackage[table]{xcolor}

\newcolumntype{d}[1]{D{.}{.}{#1}}

\begin{document}
	
\renewcommand{\arraystretch}{1.25}

\title{Gamow-Teller and double-beta decays of heavy nuclei within an effective theory}
	
\author{E.\ A.\ Coello P\'erez} 
\affiliation{Institut f\"ur Kernphysik, Technische Universit\"at Darmstadt, 64289 Darmstadt, Germany}
\affiliation{ExtreMe Matter Institute EMMI, Helmholtzzentrum f\"ur Schwerionenforschung GmbH, 64291 Darmstadt, Germany}
	
\author{J.\ Men\'endez}
\affiliation{Center for Nuclear Study, The University of Tokyo, Tokyo 113-0033, Japan}
	
\author{A.\ Schwenk} 
\affiliation{Institut f\"ur Kernphysik, Technische Universit\"at Darmstadt, 64289 Darmstadt, Germany}
\affiliation{ExtreMe Matter Institute EMMI, Helmholtzzentrum f\"ur Schwerionenforschung GmbH, 64291 Darmstadt, Germany}
\affiliation{Max-Planck-Institut f\"ur Kernphysik, Saupfercheckweg 1, 69117 Heidelberg, Germany}

\begin{abstract}

We study $\beta$ decays within an effective theory that treats
nuclei as a spherical collective core with an even number of neutrons
and protons that can couple to an additional neutron and/or proton.
First we explore Gamow-Teller $\beta$ decays of parent odd-odd
nuclei into low-lying ground, one-phonon, and two-phonon states of the
daughter even-even system. The low-energy constants of the effective
theory are adjusted to data on $\beta$ decays to ground
states or Gamow-Teller strengths. The corresponding theoretical
uncertainty is estimated based on the power counting of the effective
theory. For a variety of medium-mass and heavy isotopes the
theoretical matrix elements are in good agreement with experimental results 
within the theoretical uncertainties. We then study the two-neutrino
double-$\beta$ decay into ground and excited states. The results are
remarkably consistent with experiment within theoretical uncertainties,
without the necessity to adjust any low-energy constants.

\end{abstract}
	
	\maketitle
	
	\section{Introduction}
	
	Atomic nuclei are sensitive to fundamental interactions beyond
        the strong force that binds them.  Excited states typically decay
        due to electromagnetic interactions emitting $\gamma$ rays, while
        unstable nuclear ground states decay via
        weak interactions emitting or capturing electrons,
        neutrinos, or their antiparticles.  Nuclei are also
        used as laboratories due to their
        sensitivity to new-physics interactions beyond the Standard
        Model~\cite{severijns2011,baudis2012,engel2013}.
	
	The weak interaction is closely connected to ground-state
        decays.  Almost every unstable isotope lighter than $^{208}$Pb
        decays either via $\beta$ decay or electron
        capture. The associated half-lives can range from milliseconds
        to billions of years, with the corresponding nuclear transition
        matrix elements varying by three or more
        orders of magnitude. This wide range makes theoretical
        predictions of $\beta$ decay and electron capture particularly
        challenging tests of nuclear-structure calculations. Reliable
        predictions are also especially important for astrophysics
        because $\beta$-decay half-lives of experimentally
        inaccessible very neutron-rich nuclei set the scale of the
        rapid-neutron capture, or $r$-process, which is responsible
        for the nucleosynthesis of heavy
        elements~\cite{langanke2002,arnould2007,arcones2012}.

	{\it Ab initio} calculations of $\beta$ decays
	are still limited to few-nucleon systems (see, e.g., Refs.~\cite{gazit2009,baroni2016,klos2017})
	or focus on selected lighter isotopes~\cite{maris2011,ekstrom2014}.
	For medium-mass and heavy nuclei theoretical studies
	typically use the quasiparticle random-phase approximation
	(QRPA) (see, e.g., Refs.~\cite{niksic2005,marketin2007,niu2012,niu2013,fang2013,mustonen2014,shafer2016}),
	sometimes combined with more macroscopic calculations~\cite{moller2003},
	and when possible
	the nuclear shell model (see, e.g., Refs.~\cite{caurier2004,suzuki2011,zhi2013,kumar2016}).
	The agreement of these many-body calculations with
	experimental data demands fitting part of the nuclear interactions
	and/or the effective operators,
	such as the isoscalar pairing for the QRPA
	or a renormalization (``quenching") factor for the transition operator
	in shell model calculations~\cite{brown1988,martinezpinedo1996}.
	At present, the predictions made by different methods for non-measured decays
	disagree by factors of a few units. 
	Furthermore, in these phenomenological calculations it is difficult to provide
	well founded estimates of the associated theoretical uncertainties.
	
	Second-order processes in the weak interaction,
	double-$\beta$ ($\beta\beta$) decays,
	have been observed in otherwise stable nuclei.
	They exhibit the longest half-lives measured to date,
	exceeding $10^{19}$ years~\cite{barabash2015}.
	This decay mode is via two-neutrino $\beta\beta$ ($2\nu\beta\beta$) decay,
	to distinguish it from an even more rare type of $\beta\beta$ decay
	without neutrino emission, neutrinoless $\beta\beta$ ($0\nu\beta\beta$) decay,
	which is so far unobserved.
	The latter process is not allowed by the Standard Model
	since it violates lepton number conservation,
	and can only occur if neutrinos are their own antiparticles.
	$0\nu\beta\beta$ decay is the object of very intense
	experimental searches~\cite{GERDA17,KamLAND-Zen16,CUORE15,EXO14}
	with the goal of elucidating the nature of neutrinos.
	The calculation of matrix elements for $\beta\beta$ decays
	is specially subtle~\cite{caurier2011,vergados2012,suhonen2012,engel2017}.
	It faces challenges similar to those of $\beta$ decays,
	with the additional difficulty that $\beta\beta$ decays are suppressed.
	As with $\beta$ decays, calculations by different many-body approaches
	of unknown nuclear matrix elements vary by factors of a few units. 
	The prediction of reliable matrix elements
	with theoretical uncertainties is especially pressing
	for the interpretation and planning of present and future
	$0\nu\beta\beta$ decay experiments.
	
	The goal of this work is to use an effective theory (ET) framework
	to study the $\beta$, electron capture, and $\beta\beta$ decays
	of medium-mass and heavy nuclei.
	Calculations within an ET can provide transition matrix elements
	with quantified theoretical uncertainties,
	and are therefore a good complement to existing many-body $\beta$-decay studies.
	We focus on Gamow-Teller (GT) transitions for which experimental data are available,
	leaving the study of $0\nu\beta\beta$ decay for future work.
	The ET used here is valid for transitions involving spherical systems,
	because nuclei are treated as a spherical collective core
	coupled to a few nucleons.
	
	Over the past decades, ETs have been applied to describe
	the low-energy properties of nuclei.
	Effective theories exploit the separation of scales
	between the low-energy physics governing the processes of interest,
	which are treated explicitly,
	and the high-energy physics
	whose effects are integrated out and encoded into low-energy constants (LECs)
	that must be fit to experimental data.
	The ET is formulated
	in terms of the low-energy degrees of freedom (DOF) and their interactions,
	consistent with the symmetries of the underlying theory.
	The ET offers a systematic order-by-order expansion
	based on a power counting given by the ratio
	of the low-energy over the breakdown scale,
	at which the ET is no longer valid.
	This also allows for the quantification
	of the theoretical uncertainties at a given order in the ET~\cite{dobaczewski2014,furnstahl2015-1, furnstahl2015-2, wesolowski2016}.
	
	Effective field theories have been very successful in describing
	two- and three-nucleon forces in terms of nucleon and pion fields~\cite{vankolck1999, bedaque2002, epelbaum2009, machleidt2011, hammer2013}.
	In combination with powerful {\it ab initio} methods,
	chiral interactions have been employed to calculate low-energy properties
	of light and medium-mass isotopes (see, e.g., recent reviews~\cite{barrett2013, hagen2014,hebeler2015,navratil2016,hergert2016}).
	Complementary, a different set of ETs has been proposed
	to describe heavy nuclei in terms of collective DOF~\cite{papenbrock2011, zhang2013, papenbrock2014, coelloperez2015-1, papenbrock2015, coelloperez2015-2, coelloperez2016}.
	In particular, the ET developed in Refs.~\cite{coelloperez2015-2,coelloperez2016}
	describes the low-energy properties of spherical even-even and
	odd-mass nuclei in terms of effective single-particle DOF coupled to a collective spherical core.
	Using this approach, the low-energy spectra and electromagnetic properties
	of even-even and odd-mass nuclei (the latter with $\sfrac{1}{2}^{-}$ ground states)
	were described consistently, in good agreement with experiment~\cite{coelloperez2016}.
	The electromagnetic strength between low-lying states,
	including magnetic dipole transitions in odd-mass systems, was predicted successfully.
	Since for $\beta$ decays the relevant physics is expected to be like
	that of magnetic dipole transitions, the ET framework offers a promising approach
	to describe $\beta$, electron capture, and $\beta\beta$ decays in nuclei.
	
	This paper is organized as follows.
	Section~\ref{theories} serves as a short summary of key elements
	from the theory of $\beta$, electron capture and $\beta\beta$ decays. 
	Section~\ref{single} begins with a brief introduction to the ET
	of spherical even-even nuclei followed by an extension of the theory
	to account for the low-lying states of odd-odd nuclei.
	Next we discuss the GT transition operator
	that enters $\beta$ decays as well as methods to fix the associated LECs.
	In Sec.~\ref{sec:results_beta} we present our ET results
	for $\beta$ decay matrix elements
	of spherical odd-odd nuclei with $1^{+}_{\rm gs}$ ground states
	into final states of even-even nuclei
	corresponding to different collective ET excitations.
	We compare the ET predictions with experimental data,
	including the estimated theoretical uncertainties.
	In Sec.~\ref{double} we test the ability of the ET to 
	consistently describe $\beta$ and $2\nu\beta\beta$ decays,
	without the necessity to adjust additional LECs.
	We conclude with a brief summary and outlook in Sec.~\ref{summary}.
	
	\section{Types of \texorpdfstring{$\boldsymbol{\beta}$}{} decays}
	\label{theories}
	
	\subsection{Single-\texorpdfstring{$\boldsymbol{\beta}$}{} decay and electron capture}

	The most common weak interaction process is $\beta$
        decay.  Here either one of the $N$ neutrons in a nucleus
        decays into a proton ($\beta^-$ decay), or one of the $Z$
        protons decays into a neutron ($\beta^+$ decay).  The total
        number of nucleons $A$ remains the same.  For the electric
        charge, lepton number, and angular momentum to be conserved, it
        is required that an electron ($e^-$) or positron ($e^+$) be
        emitted along with an electron antineutrino ($\overline{\nu}_e$) or
        neutrino ($\nu_e$):
	\begin{align}
		A(Z,N) \overset{\beta^{-}}{\longrightarrow} A(Z+1,N-1)
		+ e^{-} + \overline{\nu}_{e}\,, \\
		A(Z,N) \overset{\beta^{+}}{\longrightarrow} A(Z-1,N+1)
		+ e^{+} + \nu_{e}\,.
		\label{beta}
	\end{align}
	Alternatively, proton-rich nuclei can undergo a third process,
        an electron capture (EC), which involves the capture of an
        electron by a proton, yielding a neutron.  Again, conservation
        of energy, angular momentum, and lepton number requires an
        electron neutrino to be emitted:
	\begin{equation}
	A(Z,N) + e^{-} \overset{\rm EC}{\longrightarrow}
	A(Z-1,N+1) + \nu_{e}\,.
	\label{EC}
	\end{equation}
	
	At lowest order in the weak interaction,
	it is possible to distinguish two types of dominant, so-called allowed, transitions.
	Gamow Teller and Fermi (F) decays differ in the spin dependence
	of the associated one-body operator.
	The GT and F operators are defined as
	\begin{align}
		O_{\rm GT} &= \sum_{a=1}^A
		{\boldsymbol{\sigma}}_{a} \tau^{\pm}_{a}\,,
		\label{GT} \\
		O_{\rm F} &= \sum_{a=1}^A \tau^{\pm}_{a}\,,
		\label{F}
	\end{align}
	where $\boldsymbol{\sigma}$ denotes the spin,
	$\bf{\tau}^+$ ($\bf{\tau}^-$) is the isospin raising (lowering) operator,
	and a sum is performed over all nucleons in the nucleus.
	
	In this work, we focus on allowed GT transitions, whose
	decay rates are related to the reduced matrix elements of the GT
	operator in Eq.~(\ref{GT})
	between the corresponding initial ($i$) and final ($f$) nuclear states:
	\begin{equation}
	M_{\rm GT}=\langle f || O_{\rm GT}|| i \rangle\,.
	\end{equation}
	The decay's half-life is given by
	\begin{equation}
	\label{ft}
	\frac{1}{t_{if}} = \frac{f_{if}}{\kappa}
	\frac{g_{A}^{2}\left|M_{\rm GT}\right|^{2}}{2J_{i}+1}\,,
	\quad
	{\rm or}
	\quad
	(ft)_{if} = \frac{\kappa}{g_{A}^{2}}
	\frac{2J_{i}+1}{\left|M_{\rm GT}\right|^{2}}\,.
	\end{equation}
	Here the phase-space factor $f_{if}$ contains all the information on
	the lepton kinematics, $\kappa= 6147$~s is the $\beta$-decay constant,
	$g_{A}=1.27$ is the axial-vector coupling,
	and $J_{i}$ denotes the total angular momentum of the initial state.
	The quantity $(ft)_{if} \equiv f_{if}t_{if}$ is known as the $ft$-value
	and is directly comparable to the transition nuclear matrix element.
	
	\subsection{\texorpdfstring{$\boldsymbol{2\nu\beta\beta}$}{} decays}
	
	In the $2\nu\beta\beta$ decay, two neutrons of the parent even-even nucleus
	decay into two protons.
	Two electrons and antineutrinos are emitted as well
	due to charge and lepton-number conservation:
	\begin{equation}
	A(Z,N) \overset{2\beta^{-}}{\longrightarrow} A(Z+2,N-2) + 2e^{-}
	+ 2\overline{\nu}_{e}\,.
	\end{equation}
	Such decays have been detected for several nuclei,
	including a few cases of transitions into excited states~\cite{barabash2015}.
	At present the similar but kinematically less favored
	$\beta^+\beta^+$ and the $2\nu$ double-electron-capture (ECEC) decays
	have only been observed in geochemical measurements for $^{130}$Ba~\cite{Meshik2001, Pujol2009},
	and there is also an indication of a possible detection in $^{78}$Kr~\cite{Gavrilyuk2013}.
	
	In principle, both GT and F operators enter the description of 
	$2\nu\beta\beta$ decays.
	Nevertheless, in decays to low-lying states of the final daughter nucleus
	only the GT part is relevant. The F operator does not connect states
	with different isospin quantum numbers,
	and its strength is almost completely exhausted by
	the isobaric analog state, which lies at an excitation energy of tens of MeVs.
	The decay rate for $2\nu\beta\beta$ decay is then~\cite{engel2017}
	\begin{equation}
	\frac{1}{t^{2\nu\beta\beta}_{if}} = G^{2\nu\beta\beta}_{if}
	g_A^4 \left|M^{2\nu\beta\beta}_{\rm GT}\right|^{2}\,,
	\end{equation}
	where $G^{2\nu\beta\beta}_{if}$ is a phase-space factor,
	and the nuclear matrix element $M^{2\nu\beta\beta}_{\rm GT}$ is given by
	\begin{equation}
	M^{2\nu\beta\beta}_{\rm GT} = \sqrt{\frac{1}{s}} \sum_{n}
	\frac{\langle f || \sum_{a} {\boldsymbol{\sigma}}_{a} \tau^{+}_{a}
		|| 1_n^+ \rangle \langle 1_n^+ || \sum_{b} {\boldsymbol{\sigma}}_{b}
		\tau^{+}_{b} || i \rangle }{(D_{nf}/m_{e})^s}\,,
	\label{2nuGT}
	\end{equation}
	where the electron mass $m_{e}$
	is introduced to make the matrix element dimensionless,
	$s\equiv 1+2\delta_{2J_f}$ with $J_f=0, 2$ being the spin of the
	final state, and the sum runs over all $|1^+_n\rangle$ states of the intermediate odd-odd nucleus.
	The energy denominators $D_{nf}$ are given
	in terms of the energy of the initial ($i$), final ($f$), and intermediate ($n$) states by
	\begin{equation}
	D_{nf}= E_n - \frac{E_i-E_f}{2}\,.
	\end{equation}
	
	\section{ET for Single-\texorpdfstring{$\boldsymbol{\beta}$}{} decay}
	\label{single}
	
	In this section, we formulate an ET for the GT decays of parent
	odd-odd into daughter even-even nuclei. Our approach is valid for
	spherical systems with low-energy spectra and electromagnetic
    transitions well reproduced by an ET written in terms of collective
    DOF, which at leading order (LO) represent a five-dimensional harmonic oscillator. 
The ET DOF are therefore similar to those in the collective Hamiltonian of Bohr and Mottelson~\cite{bohr1952, bohr1953-3, bohr1975, rowe2010} or the interacting boson
model~\cite{arima1975, arima1976-1, arima1976-2, otsuka1978-1, otsuka1978-2, iachello1979}. An advantage is that in the ET
    the theoretical uncertainties due to omitted
    DOF can be propagated to the nuclear matrix elements
    and decay half-lives, allowing for a more informed comparison of the ET predictions
    with experimental data.
	
	\subsection{ET for even-even and odd-odd nuclei}
	
	The ET developed in Refs.~\cite{coelloperez2015-2, 
	coelloperez2016}
	describes the low-energy properties of spherical even-even and
	odd-mass nuclei in terms of collective excitations that can be coupled
	to an odd neutron, neutron-hole, proton, or proton-hole. The effective operators are written in
	terms of creation and annihilation operators, which are the DOF of the
	ET. These include the following:
	\begin{enumerate}[i)]
		\item
		Collective phonon operators $d^{\dagger}_{\mu}$ and $d_{\mu}$,
		which create and annihilate quadrupole phonons associated with
		low-energy quadrupole excitations of the even-even core.
		\item
		Neutron operators $n^{\dagger}_{\mu}$ and $n_{\mu}$, which create
		and annihilate a neutron or neutron-hole in a $j_{n}^{\pi_{n}}$
		single-particle orbital with total angular momentum $j$ and parity $\pi$.
		\item
		Proton operators $p^{\dagger}_{\mu}$ and $p_{\mu}$, which create
		and annihilate a proton or proton-hole in a $j_{p}^{\pi_{p}}$ orbital.
	\end{enumerate}
	
	Whether the fermion operators represent a particle or a hole
	depends on the odd-mass nucleus we want to describe and the
	even-even nucleus chosen as a core. In
	Ref.~\cite{coelloperez2016}, silver isotopes with
	$\sfrac{1}{2}^{-}$ ground states were described both
	as an odd proton coupled to palladium cores and as an odd
	proton-hole coupled to cadmium cores. Both descriptions turned
	out to be consistent with each other. The above operators
	fulfill the following relations
	\begin{equation}
	\left[d_{\mu}, d^{\dagger}_{\nu} \right] = \delta_{\mu\nu}\,,
	\quad
	\left\{n_{\mu}, n^{\dagger}_{\nu} \right\} = \delta_{\mu\nu}\,,
	\quad
	\left\{p_{\mu}, p^{\dagger}_{\nu} \right\} = \delta_{\mu\nu}\,.
	\label{dof}
	\end{equation}
	While the creation operators are the components of spherical tensors,
	the annihilation operators are not. To facilitate the construction of
	spherical-tensor operators with definite ranks, we define annihilation
	spherical tensors with components
	$\tilde{a}_{\mu}=(-1)^{j_{a}+\mu}a_{-\mu}$, where $a=d,n,p$ and
	$j_{d}=2$.
	
The Hamiltonian employed in previous work to describe
the energy spectra of a particular even-even nucleus and an adjacent
odd-mass nucleus at next-to-leading order (NLO) is
\begin{equation}
\begin{split}
H_{\rm ET}^{\rm NLO} = & \omega \left(d^{\dagger} \cdot \tilde{d} \right)
+ \sum_{fl} g_{fl}
\left(d^{\dagger} \otimes \tilde{d} \right)^{(l)}
\cdot \left(f^{\dagger} \otimes \tilde{f} \right)^{(l)},
\end{split}
\label{ham}
\end{equation}where $f$ can be either $n$ or $p$, depending in
which odd-mass nucleus we want to describe, and $\omega$
and $g_{fl}$ are LECs that must be fitted to data.
The LEC accompanying the LO term, $\omega$, may
be thought of as the energy of the collective mode. It scales as the
excitation energy of the first excited $2^{+}$ state of the
even-even nucleus of interest. Terms
proportional to $g_{fl}$ are required to describe the spectrum of
the odd-mass nucleus at NLO. The energy scale $\Lambda$ at
	which the ET breaks down lies around the three-phonon level. Based on
	previous work~\cite{coelloperez2015-2}, we will set
	$\Lambda=3\omega$ in what follows, even though the breakdown
	scale might not be exactly the same for every pair of nuclei studied
	in this work. The effective operators of the theory are constructed
	order-by-order adding all relevant terms that correct the previous
	one by a positive power of $\varepsilon\equiv\omega/\Lambda$.
	
	The reference state $|0\rangle$ of the ET represents the $0^{+}_{\rm gs}$
	ground state of the even-even nucleus of interest. Multiphonon
	excitations of this state represent excited states in the even-even
	system. Of particular relevance for our work are one- and
	two-phonon excitations:
	\begin{equation}
	|2M1\rangle = d^{\dagger}_{M}|0\rangle\,,
	\quad
	{\rm and}
	\quad
	|JM2\rangle = \sqrt{\frac{1}{2}}
	\left(d^{\dagger}\otimes d^{\dagger}\right)^{(J)}_{M}|0\rangle\,,
	\label{eell}
	\end{equation}
	where in the notation $|JM\mathcal{N}\rangle$, $J$ and $M$ are
	the total angular momenta of the state and its projection,
	and $\mathcal{N}$ is the number of phonons. We define the coupling
	of two spherical tensors as in Ref~\cite{varshalovich1988},
	and refer to Ref.~\cite{rowe2010}
	for a detailed description of the construction of multi-phonon excitations.
	We highlight that the ET introduced above
	reproduces the low-lying spectra and electromagnetic
	moments and transitions of vibrational medium-mass and heavy nuclei
	within the estimated theoretical uncertainties~\cite{coelloperez2015-2}.
	
	In a similar fashion, the ground states of adjacent odd-mass nuclei
	can be described as fermion excitations of the reference state
\begin{equation}
|j_fM\rangle = f^{\dagger}_M|0\rangle\,,
\end{equation}
while excited states in these systems are represented by their 
multiphonon excitations. Even though several single-particle orbitals may be relevant to give a full description of the
odd-mass states,
	it is assumed that at LO in the ET only one orbital is required
	to describe the low-energy properties of these systems.
	The relevant single-particle orbital is inferred from
	the quantum numbers of the ground state of the odd-mass
	nucleus of interest.
	This assumption works well for odd-mass nuclei near shell closures
	with $\sfrac{1}{2}^{-}$ ground states~\cite{coelloperez2016}.
	In these systems, a reasonable agreement was found
	between the ET predictions and experimental data,
	regarding not only low-energy excitations
	but also electric and magnetic moments and transitions~\cite{coelloperez2016}.
	
	In order to describe the allowed GT $\beta$ decays of parent odd-odd nuclei,
	we extend the collective ET of Refs.~\cite{coelloperez2015-2,coelloperez2016}
	and write the low-lying positive-parity states in the odd-odd nucleus as
	\begin{equation}
	|JM; j_{p};j_{n}\rangle =
	\left(n^{\dagger}\otimes p^{\dagger}\right)^{(J)}_{M} |0 \rangle\,,
	\label{ooll}
	\end{equation}
	where the fermion operators represent particles or holes
	depending on the odd-odd nucleus of interest. For example,
	$^{80}{\rm Br}$ can be described coupling a neutron and a
	proton hole to a $^{80}{\rm Kr}$ core, or coupling a
	neutron hole and a proton to a $^{80}{\rm Se}$ core.
	The $j_{n}$ and $j_{p}$ labels in the odd-odd state indicate
	the coupling
	of the odd neutron and odd proton on top of the collective spherical ground state.
	The angular momentum and parity of the single-particle orbitals
	to be used are inferred from
	the quantum numbers of the low-lying states of the adjacent odd-mass nuclei.
	Therefore, the total angular momenta and parities
	of these orbitals must fulfill the relations
	$|j_{n}-j_{p}| \leqslant J \leqslant j_{n}+j_{p}$,
	and $\pi_{n}\pi_{p}=1$ for positive-parity states.
The correction
\begin{equation}
\begin{split}
\Delta H &{}_{\rm ET}^{\rm NLO}
= \sum_l \epsilon_l
\left(n^{\dagger} \otimes \tilde{n} \right)^{(l)} \cdot
\left(p^{\dagger} \otimes \tilde{p} \right)^{(l)},
\end{split}
\end{equation}
must be added to the Hamiltonian in Eq.~(\ref{ham}) in order to account
for the mass difference between the even-even and odd-odd
ground states. It is important to note that the LO calculation
	of single-$\beta$ decays of the ground states of odd-odd nuclei
	require us to construct their energy spectra only at LO,
	simplifying the calculations considerably as the
terms proportional to $g_{fl}$ in Eq.~(\ref{ham}) do not enter at
this order.
	Contributions due to additional DOF relevant for excited states
	are taken into account in the uncertainty estimates associated with the LO results.
	We stress that these uncertainties must be tested whenever data are available,
	since they probe the validity of the power counting
	and the reliability of the LO calculations.
	
	\subsection{Effective GT operator}
	
	Next we construct the operator corresponding to the GT operator in Eq.~(\ref{GT}),
	in terms of the effective DOF.
	We write the most general positive-parity
	spherical-tensor operator of rank one
	capable of coupling the low-lying states of the parent odd-odd nucleus
	introduced in Eq.~(\ref{ooll})
	to the ground, one-phonon, and two-phonon states of the daughter even-even nucleus
	represented in Eq.~(\ref{eell}).
	At lowest order in the number of $d$ operators, this operator 
	is given by
	\begin{align}
	\label{GTEFT}
	O_{\rm GT} &= C_{\beta}
	\left(\tilde{p} \otimes \tilde{n} \right)^{(1)} \nonumber \\
	&+ \sum\limits_{\ell} C_{\beta \ell} \left[
	\left(d^{\dagger} + \tilde{d}\right) \otimes
	\left( \tilde{p} \otimes \tilde{n} \right)^{(\ell)} \right]^{(1)} \nonumber \\
	&+ \sum\limits_{L\ell} C_{\beta L\ell} \left[
	\left(d^{\dagger} \otimes d^{\dagger} +
	\tilde{d}\otimes\tilde{d}\right)^{(L)} \otimes
	\left( \tilde{p} \otimes \tilde{n} \right)^{(\ell)} \right]^{(1)} \,,
	\end{align}
	where $C_{\beta}$, $C_{\beta \ell}$, and $C_{\beta L \ell}$ are
	LECs that must be fit to experimental data.
	
	Let us discuss the effective operator of Eq.~(\ref{GTEFT}) in more detail.
	For the allowed GT $\beta^{-}$ decay of the odd-odd nucleus with
	$N+1$ neutrons and $Z-1$ protons
	into the even-even nucleus with $N$ neutrons and $Z$ protons,
	the odd-odd system is described within the ET
	as an even-even core coupled to a neutron and a proton hole.
	For the decay to take place,
	the fermion annihilation operators
	must annihilate the odd neutron and proton hole.
	This action represents the decay of a neutron in the odd-odd system
	into a proton, which then fills the proton hole, yielding the even-even system.
	An additional step in which the odd neutron fills a neutron hole
	takes place if the annihilated neutron is part of the core;
	however, the ET cannot differentiate between the two processes.
	For the description of the $\beta^{+}$ or EC decay
	to the even-even nucleus with $N+2$ neutrons and $Z-2$ protons,
	it is more convenient to describe the odd-odd nucleus
	in terms of the later $(N+2,Z-2)$ even-even core
	coupled to a neutron hole and a proton.
	In this case, the fermion annihilation operators
	annihilate the odd neutron hole and proton,
	representing the conversion of a proton into a neutron that fills the neutron hole.
	Again, the additional filling of a proton hole by the odd proton
	follows if the annihilated proton is in the core.
	
	The reduced matrix elements of the effective GT operator in Eq.~(\ref{GTEFT})
	between low-lying states of the parent odd-odd system in Eq.~(\ref{ooll})
	and the daughter even-even ground, one-phonon, and two-phonon states in Eq.~(\ref{eell}) are
	\begin{widetext}
		\begin{align}
			M_{\rm GT}\left(J_{i}^{+}\rightarrow 0^{+}_{\rm gs}\right)
			&= \left\{\begin{array}{c l}
				-C_{\beta} \sqrt{3} (-1)^{j_{p}-j_{n}+J_{i}} &
				\quad J_{i}=1 \\
				0 & \quad {\rm otherwise}
			\end{array} \right.,
			\label{lltogs} \\
			M_{\rm GT}\left(J_{i}^{+}\rightarrow 2^{+}_{1{\rm ph}}\right)
			&= \left\{\begin{array}{c l}
				C_{\beta J_{i}} \sqrt{3} (-1)^{j_{p}-j_{n}+J_{i}} &
				\quad |J_{i}-1| \leqslant 2 \leqslant J_{i}+1 \\
				0 & \quad {\rm otherwise}
			\end{array} \right.,
			\label{llto1ph} \\
			M_{\rm GT}\left(J_{i}^{+}\rightarrow J_{2{\rm ph}}^{+} \right)
			&= \left\{\begin{array}{c l}
				C_{\beta J_{2{\rm ph}}J_{i}} \sqrt{6} (-1)^{j_{p}-j_{n}+J_{i}} &
				\quad |J_{i}-1| \leqslant J_{2{\rm ph}} \leqslant J_{i}+1 \\
				0 & \quad {\rm otherwise}
			\end{array} \right.,
			\label{llto2ph}
		\end{align}
	\end{widetext}
	where the subscripts $\rm gs$ and $n{\rm ph}$ identify
	the ground and $n$-phonon states of the daughter even-even nucleus,
	respectively. We also note that the LECs in Eqs.~(\ref{lltogs})--(\ref{llto2ph})
	implicitly take into account additional corrections
	to the GT operator in Eq.~(\ref{GT}), such as a possible ``quenching".
	
	\subsection{ET GT decay to ground and excited states}
	
\begin{figure}[b]
		\centering
		\includegraphics[width=0.8\columnwidth]{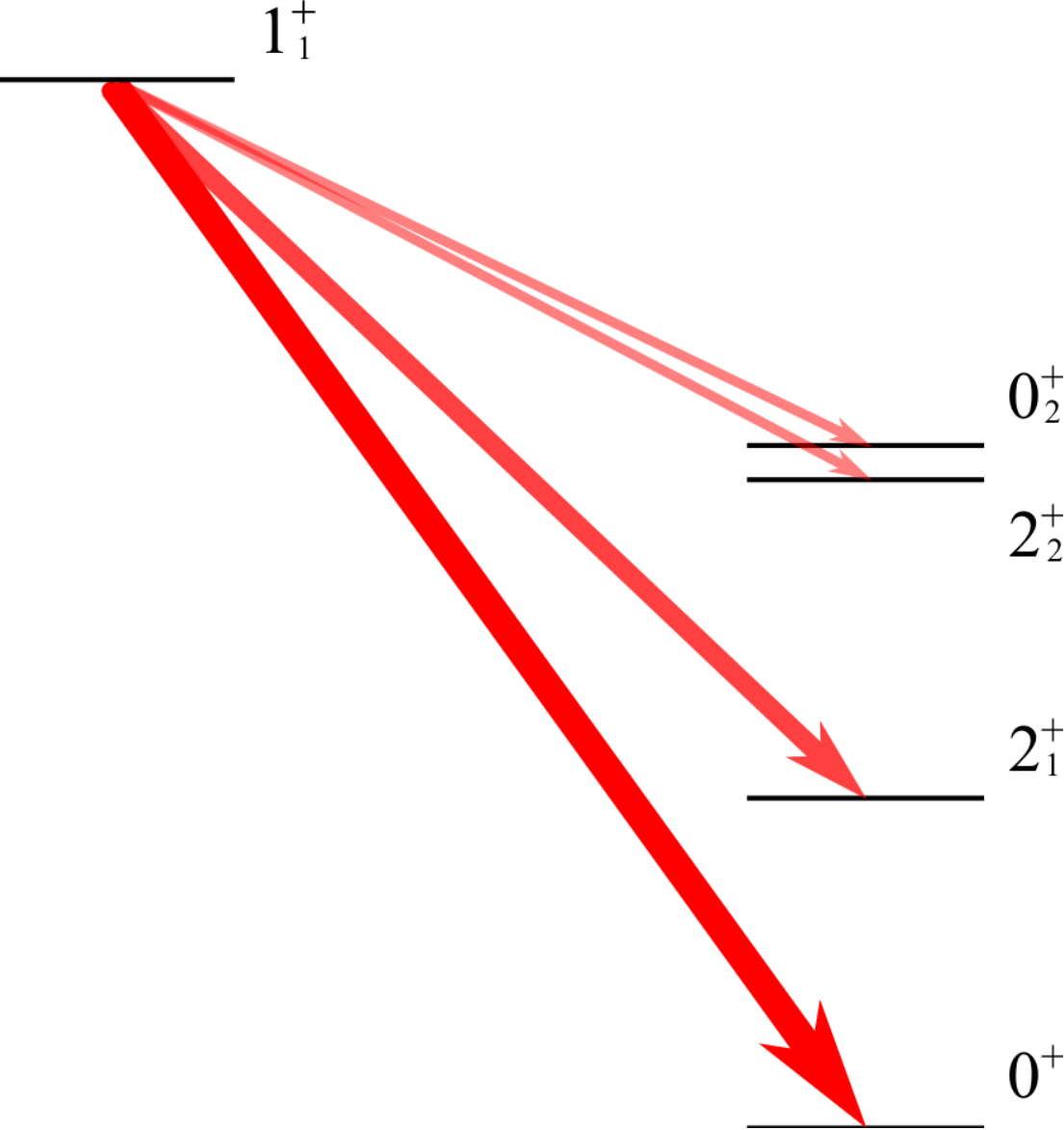}
		\caption{Schematic representation of the relative size of the
			matrix elements for the $\beta$ decays of $^{80}{\rm Br}$
			into the ground ($0^+_{\rm gs}$), one-phonon ($2^+_1$),
			and two-phonon ($0^+_2$, $2^+_2$) excited states of~$^{80}{\rm Kr}$.}
		\label{80br-80kr}
	\end{figure}

	The first, second, and third terms
	of the effective GT operator in Eq.~(\ref{GTEFT})
	couple states with phonon-number differences of zero, one, and two, respectively.
	Thus, they describe the $\beta$ decays
	from the ground state of the odd-odd nucleus
	to the ground, one-phonon, and two-phonon states in the even-even nucleus, respectively.
	Figure~\ref{80br-80kr} schematically shows the case of
	the $\beta$ decays of $^{80}{\rm Br}$
	into the $0^{+}_{\rm gs}$, $2^{+}_{1}$, $2^{+}_{2}$ and $0^{+}_{2}$ states
	of $^{80}{\rm Kr}$, identified as the ground, one-phonon, and two-phonon states.
	
	The LECs $C_{\beta}$, $C_{\beta J_i}$, and $C_{\beta J_{2{\rm ph}} J_i}$ encode
	the microscopic information of the nuclei involved in the decay.
	While the value of the LECs is not predicted by the ET,
	the power counting established in previous works~\cite{coelloperez2015-2,coelloperez2016}
	suggests scaling factors between them.
	This power counting is based on the assumption that
	at the energy scale $\Lambda$ where the ET breaks down,
	the matrix elements of every term of any effective operator scale similarly.
	
	From this assumption and the
	effective Hamiltonian describing the even-even systems,
	it can be concluded that at the breakdown scale $\Lambda$
	the matrix elements of an operator
	containing $n$ powers of $d$ operators scale as~\cite{coelloperez2015-2}
	\begin{equation}
	\langle d^{n} \rangle \sim
	\left(\frac{\Lambda}{\omega}\right)^{n/2}.
	\label{power}
	\end{equation}
	For more details, we refer the reader to Ref.~\cite{coelloperez2015-2}.
	
	The power counting in Eq.~(\ref{power}) implies that at the
	breakdown scale $\Lambda$ the matrix elements of the different
	terms in the effective GT operator~(\ref{GTEFT}) scale as
	\begin{equation}
	C_{\beta}\langle d^{0} \rangle \sim
	C_{\beta J_i} \langle d^{1} \rangle
	\quad {\rm or} \quad
	\frac{C_{\beta J_i}}{C_{\beta}} \approx
	0.58(^{+42}_{-25}) \,,
	\label{scale}
	\end{equation}
	and
	\begin{equation}
	C_{\beta} \langle d^{0} \rangle \sim
	C_{\beta J_{2{\rm ph}} J_i} \langle d^{2} \rangle
	\quad {\rm or} \quad
	\frac{C_{\beta J_{2{\rm ph}} J_i}}{C_{\beta}} \approx
	0.33(^{+25}_{-14}) \,.
	\label{scale2}
	\end{equation}
	The resulting relative sizes of the matrix elements for the
	$^{80}{\rm Br}$ decay are also shown schematically in
	Fig.~\ref{80br-80kr}. The theoretical uncertainties for the
	above ratios have been estimated based on the expectation for
	the next-order LECs $C$ to be of natural size,
	encoded into prior distributions of the form
	\begin{align}
	\text{pr}(C|c) &= \frac{1}{\sqrt{2\pi} c}
	e^{-\frac{1}{2} \left(\frac{C-1}{c}\right)^2} \,, \\[2mm]
	\text{pr}(c) &= \frac{1}{\sqrt{2\pi}\sigma c}
	e^{-\frac{1}{2} \left(\frac{\log{c}}{\sigma}\right)^2} \,,
	\end{align}
	with $\sigma=\log(3/2)$, so that a value for a LEC in the range
	$\sqrt{\omega/\Lambda}\leq C \leq \sqrt{\Lambda/\omega}$ with $\Lambda = 3\omega$
	has an associated theoretical uncertainty
	given by an interval of its probability distribution function
	with a 68\% degree of belief.
	Finally, we emphasize that these theoretical uncertainty
	estimates must be tested by comparing the ET predictions to data.
	
	\subsection{ET GT decay to ground states and\\ theoretical uncertainties}
	
	The matrix elements of single-$\beta$ decays to the ground
	state are set by the values of the LEC $C_{\beta}$, see 
	Eq.~(\ref{GTEFT}), which are to be fitted to experimental data.
	Within the ET the uncertainty of these matrix elements
	comes from two sources:
	\begin{enumerate}[i)]
		\item
		Omitted terms in the effective GT operator that involve two
		$d$ operators
		and couple states of odd-odd and even-even nuclei
		with the same number of phonons. The matrix elements of
		these terms are expected to scale as
		\begin{equation}
		\langle 0^{+}_{\rm gs} | \Delta O_{\rm GT} |
		J^{+}_{i} \rangle \sim \frac{\omega}{\Lambda} \,
		M_{\rm GT}\left( J_{i}^{+} \rightarrow 0^{+}_{\rm gs} \right).
		\end{equation}
		\item
		Next-to-leading-order corrections
		to the ground state of odd-odd nuclei
		due to terms in the Hamiltonian that can mix states
		with phonon-number differences of one.
		These corrections are expected to scale as
		$\sqrt{\omega/\Lambda}|JM;j_{p};j_{n}\rangle$ and are
		coupled to the even-even ground state by the second term
		of the effective GT operator, which contains an additional
		$d$ operator. Therefore, the corrections to
		the matrix elements also scale as
		\begin{equation}
		\langle 0^{+}_{\rm gs} | O_{\rm GT} \Delta |
		J^{+}_{i} \rangle \sim \frac{\omega}{\Lambda} \,
		M_{\rm GT}\left( J_{i}^{+} \rightarrow 0^{+}_{\rm gs} \right).
		\end{equation}
	\end{enumerate}
	
	From here, the uncertainty estimate associated to the matrix
	element in Eq.~(\ref{lltogs}) is
	\begin{equation}
	\Delta M_{\rm GT}\left( J_{i}^{+} \rightarrow 0^{+}_{\rm gs} \right)
	\sim \frac{\omega}{\Lambda} \,
	M_{\rm GT}\left( J_{i}^{+} \rightarrow 0^{+}_{\rm gs} \right).
	\label{Deltagstogs}
	\end{equation}
	Consequently, the uncertainty associated to the $\log(ft)$
	of the decay to the ground state is estimated as the next-order
	contribution to the Taylor expansion of the logarithm of the
	argument in Eq.~(\ref{ft}) with
	$M_{\rm GT}\left( J_{i}^{+} \rightarrow 0^{+}_{\rm gs} \right)
	\approx (1 \pm \omega/\Lambda)
	M_{\rm GT}\left( J_{i}^{+} \rightarrow 0^{+}_{\rm gs} \right)$, 
	that is,
	\begin{equation}
	\Delta\log(ft)_{if} \sim
	\frac{\omega}{\Lambda}\frac{2}{\ln 10} \approx 0.29\,,
	\label{Deltalog}
	\end{equation}
	where again we have assumed that $\Lambda=3\omega$.
	This uncertainty estimate is required to
	compare theory with experiment whenever $C_{\beta}$ is
	fitted to an observable. 
	
	\section{Results for single-\texorpdfstring{$\boldsymbol{\beta}$}{} decay}
	\label{sec:results_beta}
	In this section we test the ET presented in Sec.~\ref{single} by comparing its predictions for single-$\beta$ decays of odd-odd nuclei with experimental data. The ET assumes
	spherical symmetry for the nuclei involved in the decays,
	and an agreement between its predictions and experimental data
	complements other successful ET predictions for spectra, electromagnetic transition strengths and static moments~\cite{coelloperez2015-2,coelloperez2016}.
	
	\subsection{GT decays to excited states}
	\label{GT_excited_results}
	
	We begin our calculations by studying the $\beta$ decay and electron capture
	of spherical odd-odd parent nuclei with $1^{+}_{\rm gs}$ ground states
	into different ground and excited $0^{+}$ and $2^{+}$ states
	of the even-even daughter nuclei.
	This will show to which extent the ET can describe
	processes involving individual nucleons and whether
	the transitions scale as expected in the ET.
	
	For each parent nucleus,
	the LEC $C_{\beta}$ can be fitted to the transition to the ground state.
	Then the GT decays into excited collective states are predicted by the ET
	according to the scaling factors in Eqs.~(\ref{scale}) and~(\ref{scale2}):
	\begin{align}
		\frac{\left|M_{\rm GT}\left( \rm gs \rightarrow \rm 1ph \right)\right|}
		{\left|M_{\rm GT}\left( \rm gs \rightarrow \rm gs \right)\right|} &= 
		\sqrt{\frac{(ft)_{\rm gs-gs}}{(ft)_{\rm gs-1ph}}} \nonumber \\
		&= \frac{C_{\beta 1}}{C_{\beta}}
		\approx 0.58(^{+42}_{-25})\,,
		\label{ratio2} \\
		\frac{\left|M_{\rm GT}\left( \rm gs \rightarrow \rm 2ph \right)\right|}
		{\left|M_{\rm GT}\left( \rm gs \rightarrow \rm gs \right)\right|} &=
		\sqrt{\frac{(ft)_{\rm gs-gs}}{(ft)_{\rm gs-2ph}}} \nonumber \\
		&= \frac{\sqrt{2}C_{\beta J_{2 {\rm ph}}1}}{C_{\beta}}
		\approx 0.47(^{+35}_{-20})\,.
		\label{ratio3}
	\end{align}
	Thus, the ET predicts equal half-lives for the decays
	into the even-even two-phonon states at LO. This LO result is similar to the prediction of collective models for electric
	quadrupole transition strengths from two-phonon states into the
	one-phonon state, which are predicted to be equal at low-orders. In the ET, the degeneracy is lifted at NLO.
	We also note that the ET naturally predicts
	the observed successive hindering reported in Ref.~\cite{sakai1962}
	of the matrix elements for GT $\beta$ decays
	from $1^{+}_{\rm gs}$, $2^{+}_{\rm gs}$, and $3^{+}_{\rm gs}$ ground states of odd-odd nuclei
	into $0^{+}_{\rm gs}$, $2^{+}_{1}$, and $2^{+}_{2}$ states of the even-even daughter.
	
\begin{figure*}
		\centering%
		\includegraphics[width=0.65\textwidth]{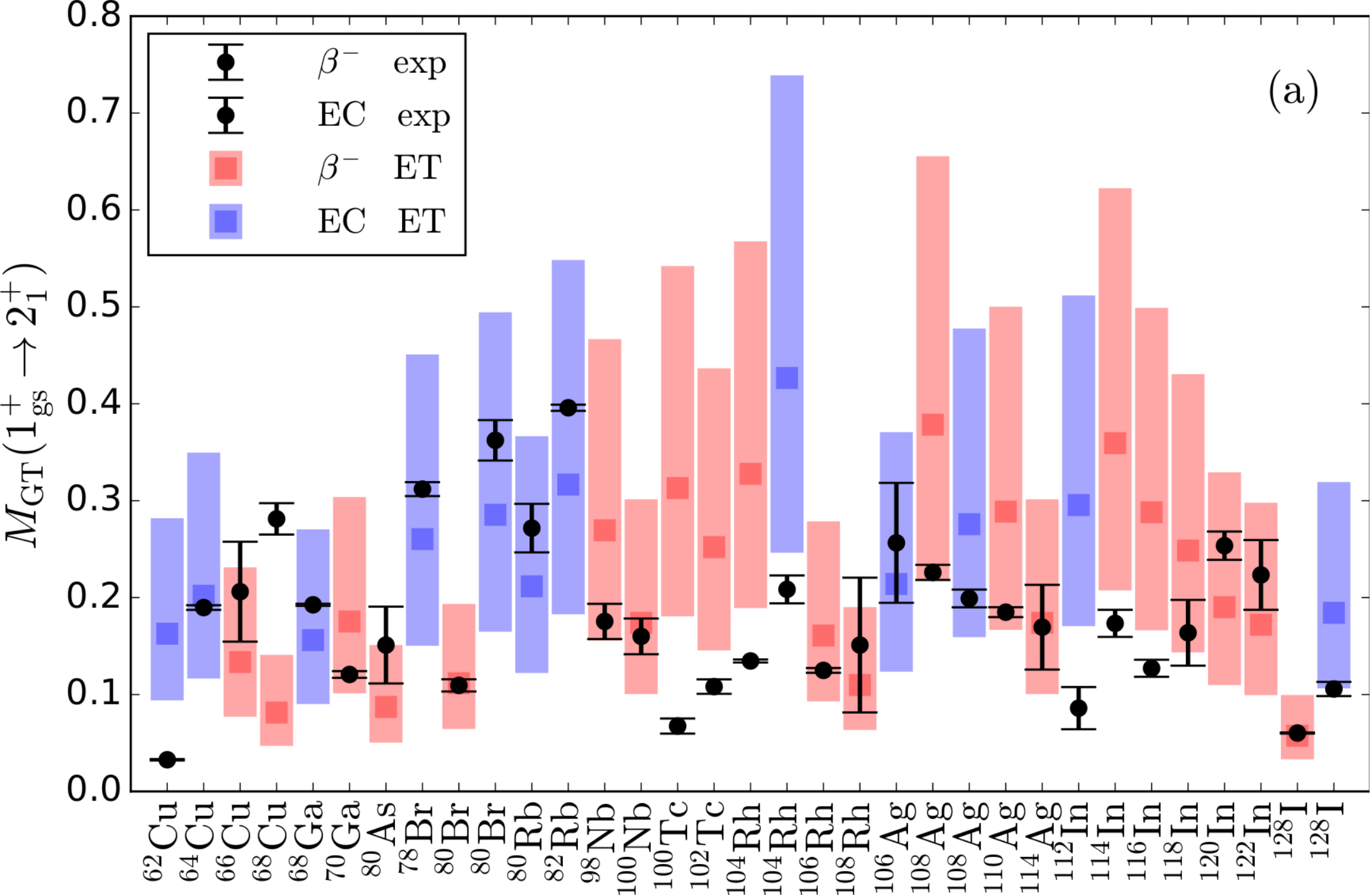}
		\includegraphics[width=0.65\textwidth]{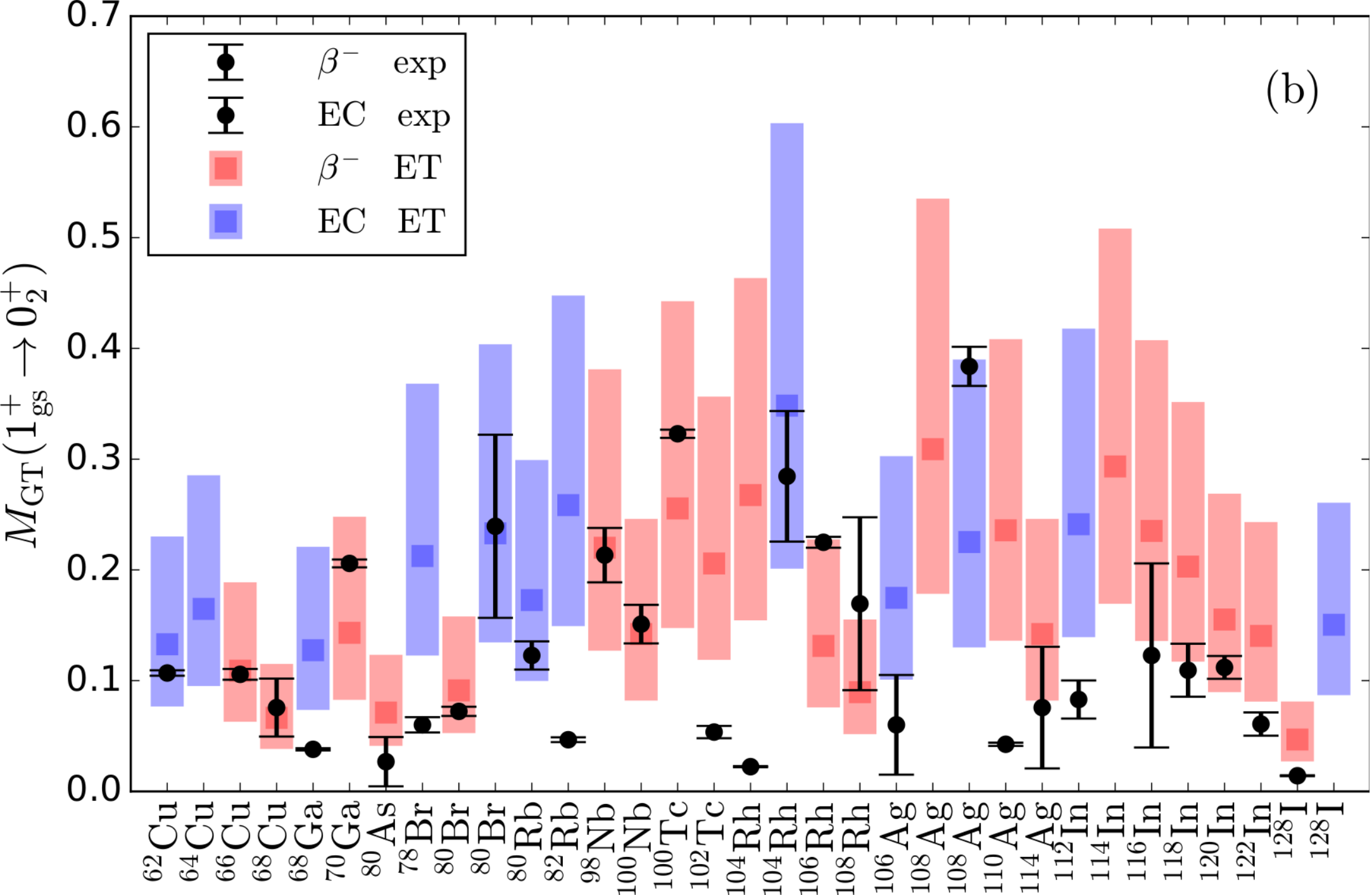}
		\includegraphics[width=0.65\textwidth]{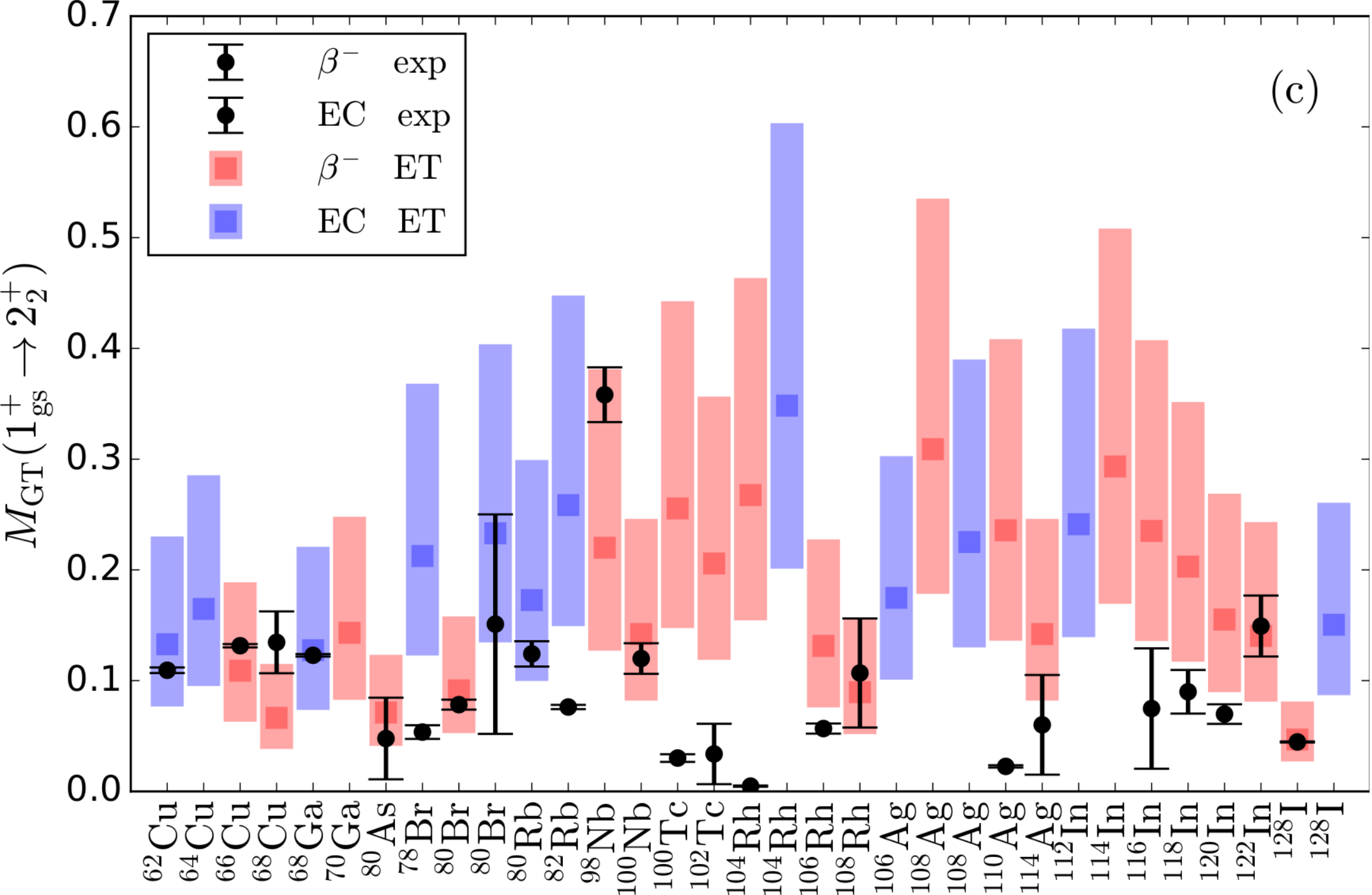}
		\caption{Calculated ET matrix elements for GT $\beta$ (red bands)
			and EC (blue bands) decays from parent odd-odd nuclei
			with $1^{+}_{\rm gs}$ ground states
			into the $2^{+}_{1}$ (a), $0^{+}_{2}$ (b),
			and $2^{+}_{2}$ (c) excited states of the daughter even-even nuclei,
			compared to experimental results (black circles) from Refs.~\cite{nichols2012,
				singh2007, browne2010-1, mccuthan2012, gurdal2016, singh2005,
				farhan2009, tuli2003, singh2003, singh2008, defrenne2009-1,
				blachot2007, defrenne2008, blachot2000, gurdal2012, lalkovski2015,
				blachot2012, blachot2010-1, kitao1995, kitao2002, tamura2007,
				katakura2008, elekes2015, singh2001}. For details, see Table~\ref{oo-eebdecay}.}
		\label{GTexcited}
	\end{figure*}

	Figure~\ref{GTexcited} compares our ET predictions with experiment
	for GT $\beta$ and EC decay matrix elements
	for a broad range of medium-mass and heavy odd-odd nuclei
	with mass numbers from $A=62$ to $A=128$.
	All parent nuclei have $1^{+}_{\rm gs}$ ground states and decay into
	excited $2^{+}_1$, $0^{+}_2$, and $2^{+}_2$ states
	of the corresponding even-even nuclei.
	Within the ET, the $2^{+}_1$ states are treated as one-phonon excitations
	of the even-even core, while the $0^{+}_2$ and $2^{+}_2$ states
	are considered to be two-phonon excitations.
	The ET results with uncertainties are calculated
	according to Eqs.~(\ref{ratio2}) and~(\ref{ratio3})
	after adjusting the LEC $C_{\beta}$ to the matrix element
	of the decay to the $0^+_{\rm gs}$ ground state of the daughter nucleus.
	The same figure shows that most of the experimental data,
	including $\beta$ GT and EC decays,
	is consistent with the ET results.
	Inconsistencies are larger for the $0^{+}_2$ and $2^{+}_2$ two-phonon states
	where the ET matrix elements tend to be overestimated,
	especially for the $\beta$ decays into $2^{+}_2$ excited states
	around mass number $A \sim 110$.
	This is not unexpected because two-phonon states lie closer
	to the ET breakdown scale.
	In Table~\ref{oo-eebdecay}, we list the values
	corresponding to all the theoretical and experimental matrix elements
	shown in Fig.~\ref{GTexcited}.
	
	\begin{table*}
		\caption{Experimental and calculated ET matrix elements
			for GT $\beta$ and EC
			decays from parent odd-odd nuclei
			with $1^{+}_{\rm gs}$ ground states (treated as a spherical collective core coupled
			to a neutron and a proton)
			into $2^{+}_1$ (one-phonon state), $0^{+}_2$, and $2^{+}_2$ (two-phonon)
			excited states of the daughter even-even nuclei. The LECs $C_\beta$ in
			the effective GT operator in Eq.~(\ref{GTEFT}) were fitted to reproduce the $\log(ft)$ values of decays to the $0^{+}_{\rm gs}$
			states, taking $g_A=1.27$.
			The experimental values were calculated from the
			$\log(ft)$-values for the transitions to the $0^{+}_{\rm gs}$, $2^{+}_1$, $0^{+}_2$, 
			and $2^{+}_2$ states,
			taken from Refs.~\cite{nichols2012,
				singh2007, browne2010-1, mccuthan2012, gurdal2016, singh2005,
				farhan2009, tuli2003, singh2003, singh2008, defrenne2009-1,
				blachot2007, defrenne2008, blachot2000, gurdal2012, lalkovski2015,
				blachot2012, blachot2010-1, kitao1995, kitao2002, tamura2007,
				katakura2008, elekes2015, singh2001}.
			The theoretical uncertainties are estimated
			assuming all LECs to be of natural size.} 
		\centering
		\begin{tabular}{c | c c | c c | c c | c c }
			\hline\hline
			${\rm Parent}\rightarrow{\rm Daughter}$ &
			\multicolumn{2}{c |}
			{$M_{\rm GT}\left( 1^{+}_{\rm gs} \rightarrow 0^{+}_{\rm gs} \right)$} &
			\multicolumn{2}{c |}
			{$M_{\rm GT}\left( 1^{+}_{\rm gs} \rightarrow 2^{+}_{1} \right)$} &
			\multicolumn{2}{c |}
			{$M_{\rm GT}\left( 1^{+}_{\rm gs} \rightarrow 0^{+}_{2} \right)$} &
			\multicolumn{2}{c}
			{$M_{\rm GT}\left( 1^{+}_{\rm gs} \rightarrow 2^{+}_{2} \right)$} \\
			& Expt. & ET & Expt. & ET & Expt. & ET & Expt. & ET \\
			\hline
			$^{62}{\rm Cu}\overset{\rm EC}{\longrightarrow}{}^{62}{\rm Ni}$ &
			$0.282(1)$ & $0.282(94)$ &
			$0.033(1)$ & $0.163(^{+119}_{-69})$ &
			$0.107(2)$ & $0.133(^{+97}_{-56})$ &
			$0.109(3)$ & $0.133(^{+97}_{-56})$  \\
			$^{64}{\rm Cu}\overset{\rm EC}{\longrightarrow}{}^{64}{\rm Ni}$ &
			$0.350(2)$ & $0.350(117)$ &
			$0.190(2)$ & $0.202(^{+148}_{-85})$ &
			& & 
			& \\
			$^{66}{\rm Cu}\overset{\beta^{-}}{\longrightarrow}{}^{66}{\rm Zn}$ &
			$0.231(14)$ & $0.231(77)$ &
			$0.206(26)$ & $0.134(^{+98}_{-56})$ &
			$0.106(5)$ & $0.109(^{+80}_{-46})$ &
			$0.132(2)$ & $0.109(^{+80}_{-46})$ \\
			$^{68}{\rm Cu}\overset{\beta^{-}}{\longrightarrow}{}^{68}{\rm Zn}$ &
			$0.141(10)$ & $0.141(47)$ &
			$0.281(16)$ & $0.081(^{+60}_{-34})$ &
			$0.076(26)$ & $0.066(^{+49}_{-28})$ &
			$0.135(28)$ & $0.066(^{+49}_{-28})$ \\
			$^{68}{\rm Ga}\overset{\rm EC}{\longrightarrow}{}^{68}{\rm Zn}$ &
			$0.270(1)$ & $0.270(90)$ &
			$0.193(1)$ & $0.156(^{+114}_{-66})$ &
			$0.038(1)$ & $0.127(^{+93}_{-54})$ &
			$0.123(1)$ & $0.127(^{+93}_{-54})$ \\
			$^{70}{\rm Ga}\overset{\beta^{-}}{\longrightarrow}{}^{70}{\rm Ge}$ &
			$0.304(1)$ & $0.304(101)$ &
			$0.121(3)$ & $0.175(^{+128}_{-74})$ &
			$0.206(4)$ & $0.143(^{+105}_{-61})$ &
			& $0.143(^{+105}_{-61})$ \\
			$^{80}{\rm As}\overset{\beta^{-}}{\longrightarrow}{}^{80}{\rm Se}$ &
			$0.151(20)$ & $0.151(50)$ &
			$0.151(40)$ & $0.087(^{+64}_{-37})$ &
			$0.027(22)$ & $0.071(^{+52}_{-30})$ &
			$0.048(37)$ & $0.071(^{+52}_{-30})$ \\
			$^{78}{\rm Br}\overset{\rm EC}{\longrightarrow}{}^{78}{\rm Se}$ &
			$0.451(5)$ & $0.451(150)$ &
			$0.312(7)$ & $0.260(^{+191}_{-110})$ &
			$0.060(7)$ & $0.213(^{+156}_{-90})$ &
			$0.054(6)$ & $0.213(^{+156}_{-90})$ \\
			$^{80}{\rm Br}\overset{\beta^{-}}{\longrightarrow}{}^{80}{\rm Kr}$ &
			$0.193(1)$ & $0.193(64)$ &
			$0.109(6)$ & $0.112(^{+82}_{-47})$ &
			$0.072(4)$ & $0.091(^{+67}_{-39})$ &
			$0.078(5)$ & $0.091(^{+67}_{-39})$ \\
			$^{80}{\rm Br}\overset{\rm EC}{\longrightarrow}{}^{80}{\rm Se}$ &
			$0.494(28)$ & $0.494(165)$ &
			$0.362(21)$ & $0.285(^{+209}_{-121})$ &
			$0.239(83)$ & $0.233(^{+171}_{-99})$ &
			$0.151(50)$ & $0.233(^{+171}_{-99})$ \\
			$^{80}{\rm Rb}\overset{\rm EC}{\longrightarrow}{}^{80}{\rm Kr}$ &
			$0.367(25)$ & $0.367(122)$ &
			$0.272(25)$ & $0.212(^{+155}_{-89})$ &
			$0.123(13)$ & $0.173(^{+126}_{-73})$ &
			$0.124(11)$ & $0.173(^{+126}_{-73})$ \\
			$^{82}{\rm Rb}\overset{\rm EC}{\longrightarrow}{}^{82}{\rm Kr}$ &
			$0.548(3)$ & $0.548(183)$ &
			$0.396(3)$ & $0.317(^{+232}_{-134})$ &
			$0.047(2)$ & $0.259(^{+189}_{-109})$ &
			$0.077(2)$ & $0.259(^{+189}_{-109})$ \\
			\hline
			$^{98}{\rm Nb}\overset{\beta^{-}}{\longrightarrow}{}^{98}{\rm Mo}$ &
			$0.467(32)$ & $0.467(156)$ &
			$0.175(18)$ & $0.269(^{+197}_{-114})$ &
			$0.213(25)$ & $0.220(^{+161}_{-93})$ &
			$0.358(25)$ & $0.220(^{+161}_{-93})$ \\
			$^{100}{\rm Nb}\overset{\beta^{-}}{\longrightarrow}{}^{100}{\rm Mo}$ &
			$0.301(35)$ & $0.301(100)$ &
			$0.160(18)$ & $0.174(^{+127}_{-74})$ &
			$0.151(17)$ & $0.142(^{+104}_{-60})$ &
			$0.120(14)$ & $0.142(^{+104}_{-60})$ \\
			$^{100}{\rm Tc}\overset{\beta^{-}}{\longrightarrow}{}^{100}{\rm Ru}$ &
			$0.542(6)$ & $0.542(181)$ &
			$0.067(8)$ & $0.313(^{+229}_{-132})$ &
			$0.323(4)$ & $0.256(^{+187}_{-108})$ &
			$0.030(2)$ & $0.256(^{+187}_{-108})$ \\
			$^{102}{\rm Tc}\overset{\beta^{-}}{\longrightarrow}{}^{102}{\rm Ru}$ &
			$0.437(7)$ & $0.437(146)$ &
			$0.108(7)$ & $0.252(^{+185}_{-107})$ &
			$0.054(6)$ & $0.206(^{+151}_{-87})$ &
			$0.034(14)$ & $0.206(^{+151}_{-87})$ \\
			$^{104}{\rm Rh}\overset{\beta^{-}}{\longrightarrow}{}^{104}{\rm Pd}$ &
			$0.568(7)$ & $0.568(189)$ &
			$0.135(2)$ & $0.328(^{+240}_{-139})$ &
			$0.022(1)$ & $0.268(^{+196}_{-113})$ &
			$0.005(1)$ & $0.268(^{+196}_{-113})$ \\
			$^{104}{\rm Rh}\overset{\rm EC}{\longrightarrow}{}^{104}{\rm Ru}$ &
			$0.739(1)$ & $0.739(246)$ &
			$0.208(14)$ & $0.427(^{+312}_{-180})$ &
			$0.285(59)$ & $0.348(^{+255}_{-147})$ &
			& $0.348(^{+225}_{-147})$ \\
			$^{106}{\rm Rh}\overset{\beta^{-}}{\longrightarrow}{}^{106}{\rm Pd}$ &
			$0.279(2)$ & $0.279(93)$ &
			$0.125(2)$ & $0.161(^{+118}_{-68})$ &
			$0.225(5)$ & $0.131(^{+96}_{-56})$ &
			$0.057(5)$ & $0.131(^{+96}_{-56})$ \\
			$^{108}{\rm Rh}\overset{\beta^{-}}{\longrightarrow}{}^{108}{\rm Pd}$ &
			$0.190(7)$ & $0.190(63)$ &
			$0.151(70)$ & $0.110(^{+80}_{-46})$ &
			$0.169(78)$ & $0.090(^{+66}_{-38})$ &
			$0.107(49)$ & $0.090(^{+66}_{-38})$ \\
			$^{106}{\rm Ag}\overset{\rm EC}{\longrightarrow}{}^{106}{\rm Pd}$ &
			$0.371(21)$ & $0.371(124)$ &
			$0.257(31)$ & $0.214(^{+157}_{-90})$ &
			$0.060(22)$ & $0.175(^{+128}_{-74})$ &
			& $0.175(^{+128}_{-74})$ \\
			$^{108}{\rm Ag}\overset{\beta^{-}}{\longrightarrow}{}^{108}{\rm Cd}$ &
			$0.656(7)$ & $0.656(219)$ &
			$0.226(8)$ & $0.378(^{+277}_{-160})$ &
			& &
			& \\
			$^{108}{\rm Ag}\overset{\rm EC}{\longrightarrow}{}^{108}{\rm Pd}$ &
			$0.478(16)$ & $0.478(159)$ &
			$0.199(9)$ & $0.276(^{+202}_{-117})$ &
			$0.384(18)$ & $0.225(^{+165}_{-95})$ &
			& $0.225(^{+165}_{-95})$ \\
			$^{110}{\rm Ag}\overset{\beta^{-}}{\longrightarrow}{}^{110}{\rm Cd}$ &
			$0.500(2)$ & $0.500(167)$ &
			$0.186(5)$ & $0.289(^{+211}_{-122})$ &
			$0.043(1)$ & $0.236(^{+173}_{-100})$ &
			$0.023(1)$ & $0.236(^{+173}_{-100})$ \\
			%
			%
			$^{114}{\rm Ag}\overset{\beta^{-}}{\longrightarrow}{}^{114}{\rm Cd}$ &
			$0.301(18)$ & $0.301(100)$ &
			$0.169(22)$ & $0.174(^{+127}_{-74})$ &
			$0.076(27)$ & $0.142(^{+104}_{-60})$ &
			$0.060(22)$ & $0.142(^{+104}_{-60})$ \\
			$^{112}{\rm In}\overset{\rm EC}{\longrightarrow}{}^{112}{\rm Cd}$ &
			$0.512(35)$ & $0.512(171)$ &
			$0.086(22)$ & $0.295(^{+216}_{-125})$ &
			$0.083(17)$ & $0.241(^{+177}_{-102})$ &
			& $0.241(^{+177}_{-102})$ \\
			$^{114}{\rm In}\overset{\beta^{-}}{\longrightarrow}{}^{114}{\rm Sn}$ &
			$0.622(1)$ & $0.622(207)$ &
			$0.173(14)$ & $0.359(^{+263}_{-152})$ &
			& &
			& \\
			%
			%
			$^{116}{\rm In}\overset{\beta^{-}}{\longrightarrow}{}^{116}{\rm Sn}$ &
			$0.499	(3)$ & $0.499(166)$ &
			$0.127(9)$ & $0.288(^{+211}_{-122})$ &
			$0.123(42)$ & $0.235(^{+172}_{-99})$ &
			$0.075(27)$ & $0.235(^{+172}_{-99})$ \\
			$^{118}{\rm In}\overset{\beta^{-}}{\longrightarrow}{}^{118}{\rm Sn}$ &
			$0.431(15)$ & $0.431(144)$ &
			$0.164(34)$ & $0.249(^{+182}_{-105})$ &
			$0.109(24)$ & $0.203(^{+149}_{-86})$ &
			$0.090(20)$ & $0.203(^{+149}_{-86})$ \\
			$^{120}{\rm In}\overset{\beta^{-}}{\longrightarrow}{}^{120}{\rm Sn}$ &
			$0.329(8)$ & $0.329(110)$ &
			$0.254(15)$ & $0.190(^{+139}_{-80})$ &
			$0.112(10)$ & $0.155(^{+114}_{-66})$ &
			$0.070(9)$ & $0.155(^{+114}_{-66})$ \\
			$^{122}{\rm In}\overset{\beta^{-}}{\longrightarrow}{}^{122}{\rm Sn}$ &
			$0.298(34)$ & $0.298(99)$ &
			$0.223(36)$ & $0.172(^{+126}_{-73})$ &
			$0.061(11)$ & $0.140(^{+103}_{-59})$ &
			$0.149(28)$ & $0.140(^{+103}_{-59})$ \\
			\hline
			$^{128}{\rm I}\overset{\beta^{-}}{\longrightarrow}{}^{128}{\rm Xe}$ &
			$0.099(1)$ & $0.099(33)$ &
			$0.060(1)$ & $0.057(^{+42}_{-24})$ &
			$0.014(1)$ & $0.047(^{+34}_{-20})$ &
			$0.045(1)$ & $0.047(^{+34}_{-20})$ \\
			$^{128}{\rm I}\overset{\rm EC}{\longrightarrow}{}^{128}{\rm Te}$&
			$0.319(18)$ & $0.319(106)$ &
			$0.106(7)$ & $0.184(^{+135}_{-78})$ &
			& &
			& \\
			\hline\hline
		\end{tabular}
		\label{oo-eebdecay}
	\end{table*}
	
	\subsection{GT decays to ground states using GT transition strengths}
	
	In Sec.~\ref{GT_excited_results},
	we have used experimental data on single-$\beta$ decays to ground states
	to fit the value of $C_{\beta}$ and then predicted the matrix elements
	for transitions to excited states of the same nucleus. Next, we study
	whether it is possible to employ other data to fit the LECs
	and in turn predict the $\beta$-decay matrix elements to ground states.
	
	Besides weak processes,
	GT strengths studied in charge-exchange reactions (via the strong interaction)
	are also sensitive to the GT spin-isospin operator,
	because the zero-degree differential cross section
	of the reaction is proportional to the GT strength~\cite{watson1985}
	\begin{equation}
	S_{\pm}(i\rightarrow f)=|\langle f ||
	{\boldsymbol{\sigma}}\tau^{\pm} || i \rangle|^{2} \,.
	\label{GTstrength}
	\end{equation}
	Therefore, reactions such as $(p,n)$ or $(^{3}{\rm He},t)$
	have the same form as $\beta^{-}$ decays,
	while $(n,p)$ reactions are related to $\beta^{+}$-like transitions.
	The GT strengths and $\beta$ decays of isospin-mirror nuclei
	have been found to be consistent in medium-mass systems~\cite{molina2015}.
	
\begin{figure}[b]
		\centering
		\includegraphics[width=1.0\columnwidth]{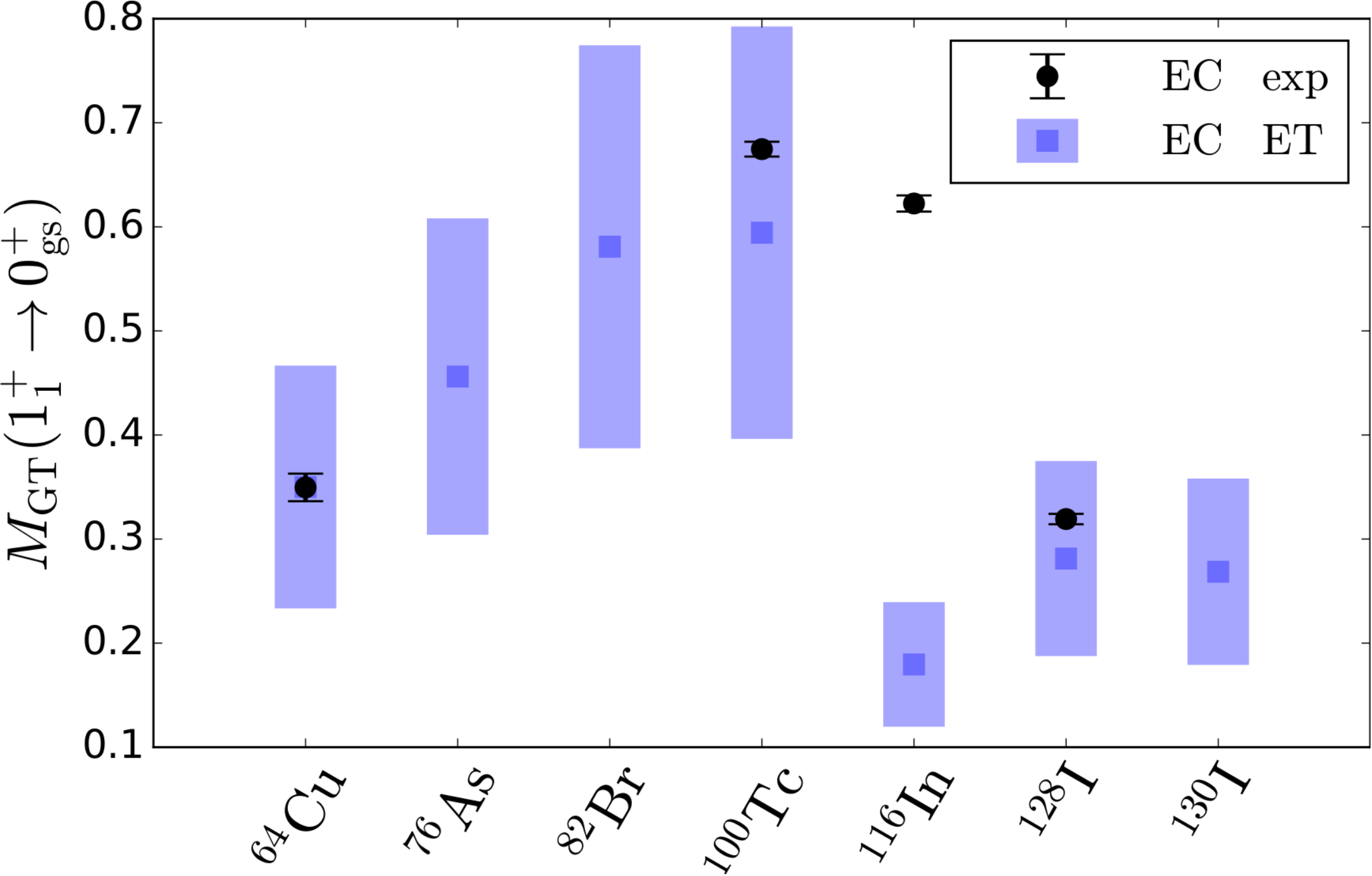}
		\caption{Calculated GT matrix elements for the transition from the
		$1^+_1$ states of odd-odd nuclei to the $0^{+}_{\rm gs}$ ground states
		of even-even nuclei, using as ET input the GT transition strengths
		measured in $(^3{\rm He},t)$ charge-exchange reactions (blue bands),
		see also Table~\ref{pGTstrength}. The ET
		results are compared to experiment calculated from the $\log(ft)$-values
		of the corresponding EC decays (black circles) from Refs.~\cite{singh2007, singh2008, blachot2010-1, elekes2015}.} 
		\label{CE}
	\end{figure}

	In this spirit, we can use the GT transition strengths
	measured in $(^3{\rm He},t)$ charge-exchange reactions
	to fit the LECs of the effective GT operator in Eq.~(\ref{GTEFT}).
	In this way, we can predict the GT matrix elements
	for the transition to the $0^{+}_{\rm gs}$ ground states
	of spherical even-even nuclei.
	Figure~\ref{CE} shows that the ET results for the GT matrix elements
	including the theoretical uncertainty,
	agree very well with experiment in three of the four cases where
	data are available. The experimental values are calculated from the $\log(ft)$-values
	of the corresponding EC decays from Refs.~\cite{singh2007, singh2008, blachot2010-1, elekes2015}. 
	The remaining results in Fig.~\ref{CE} show ET predictions for single-$\beta$ decay of
	additional $1^+$ states, which are, however, excited states of the odd-odd system
	and decay via electromagnetic transitions.

	The details of the ET predictions in Fig.~\ref{CE}
	are given in Table~\ref{pGTstrength}, which lists the
	GT strengths measured in $(^3{\rm He},t)$ charge-exchange reactions
	with initial even-even and final odd-odd nuclei for $A=64$--130~\cite{popescu2009, thies2012-1, frekers2016, thies2012-2, akimune1997, puppe2012}.
	For each reaction, Table~\ref{pGTstrength} gives
	the experimental partial GT strength to the lowest $1^+_1$ state
	of the odd-odd nucleus (the state expected to be well described by the ET).	
	An exception is the case of the
	$^{76}{\rm Ge}(^{3}{\rm He},t)^{76}{\rm As}$ reaction,
	where all the GT strength below 500~keV was taken into account
	(as reported in Table IV of Ref.~\cite{thies2012-1}),
	corresponding to three close-lying $1^{+}$ states.
	The resulting values for $C_{\beta}$ fit to the partial GT strength are given in
	Table~\ref{pGTstrength} including the comparison of the ET results to the
	experimental $\log(ft)$-values.

	\begin{table}
		\caption{Selected $(^3{\rm He},t)$ charge-exchange reactions (first column),
			experimental partial GT strengths (second column) from Refs.~\cite{popescu2009, thies2012-1, frekers2016, thies2012-2, akimune1997, puppe2012},
			and $C_{\beta}$ values fitted to them (third column).
			The fourth and fifth columns compare the
			experimental and ET results for the $\log(ft)$ values of the corresponding EC decays,
			where the experimental values are taken from Refs.~\cite{singh2007, singh2008, blachot2010-1, elekes2015}.} 
		\centering
		\begin{tabular}{ c | c c | c c }
			\hline\hline
			${\rm Reaction}$ &
			$S(0^{+}_{\rm gs}\rightarrow 1^{+}_1)$ &
			$C_{\beta}$ &
			\multicolumn{2}{c}{$\log(ft)$} \\
			& & & Expt. & ET \\
			\hline
			$^{64}{\rm Ni}(^{3}{\rm He},t){}^{64}{\rm Cu}$ & 
			$0.123$ & $0.202$ & $4.97$ & $4.97(29)$ \\
			$^{76}{\rm Ge}(^{3}{\rm He},t){}^{76}{\rm As}$\footnotemark[1] & 
			$0.210$ & $0.265$ & & $4.74(29)$ \\
			$^{82}{\rm Se}(^{3}{\rm He},t){}^{82}{\rm Br}$ & 
			$0.338$ & $0.336$ & & $4.53(29)$ \\
			\hline
			$^{100}{\rm Mo}(^{3}{\rm He},t){}^{100}{\rm Tc}$ & 
			$0.348$ & $0.341$ & $4.40$ & $4.51(29)$ \\
			$^{116}{\rm Cd}(^{3}{\rm He},t){}^{116}{\rm In}$ & 
			$0.032$ & $0.103$ & $4.47$ & $5.55(29)$ \\
			\hline
			$^{128}{\rm Te}(^{3}{\rm He},t){}^{128}{\rm I}$ & 
			$0.079$ & $0.162$ & $5.05$ & $5.16(29)$ \\
			$^{130}{\rm Te}(^{3}{\rm He},t){}^{130}{\rm I}$ & 
			$0.072$ & $0.155$ & & $5.20(29)$ \\
			\hline\hline
		\end{tabular}
		\footnotetext[1]{Comprises the sum of GT strength below 500~keV, see text.}
		\label{pGTstrength}
	\end{table}
	
	\section{ET for \texorpdfstring{$\boldsymbol{2\nu\beta\beta}$}{} decay}
	\label{double}
	
	In this section, we present the ET
	for the $2\nu\beta\beta$ decay of spherical nuclei.
	The decay calculations involve a sum over all $1^+$ states
	in the intermediate odd-odd nucleus,
	which in general are not well described by the ET.
	We overcome this limitation by
	assuming the single-state dominance (SSD) approximation,
	which requires explicitly only the lowest $1^+$ state.
	The associated uncertainty is estimated within the ET,
	and turns out to be comparable to the uncertainty of LO ET calculations.
	
	\subsection{Effective \texorpdfstring{$\boldsymbol{2\nu\beta\beta}$}{} matrix elements for decays\\ into ground and excited states}
	
	The $2\nu\beta\beta$ decay nuclear matrix element
	is given in Eq.~(\ref{2nuGT}).
	Its calculation involves a sum
	over the contributions of all $1^{+}$ states of the intermediate odd-odd nucleus. 
	Since the ET is designed to reproduce only the lowest energy states of an isotope,
	we use the SSD approximation,
	which reduces the sum to the single contribution of the lowest $1^{+}$ state. The calculation of the matrix
	 element within the closure approximation would yield a result
	 with a larger uncertainty estimate, as contributions
	 from high-lying intermediate $1^{+}$ states are not suppressed by the energy denominator. In the SSD approximation, the
	 $2\nu\beta\beta$ decay matrix element takes the form
	\begin{equation}
	\begin{split}
	M^{2\nu\beta\beta}_{\rm GT}(0^{+}_{\rm gs}\rightarrow f)
	\approx & \frac{
		M_{\rm GT}(1^{+}_{1}\rightarrow f)
		M_{\rm GT}(0^{+}_{\rm gs}\rightarrow 1^{+}_{1})}
	{\sqrt{s}(D_{1f}/m_{e})^s } \\
	= & \frac{3\kappa}{\sqrt{s}g_{A}^{2}}\bigg(\frac{m_e}{D_{1f}}\bigg)^s
	\sqrt{\frac{1}{(ft)_{1^{+}_{1}\,f}
			(ft)_{1^{+}_{1}\,0^{+}_{\rm gs}}}}\,,
	\end{split}
	\label{SSD}
	\end{equation}
	where $s\equiv 1+2\delta_{2J_f}$, and we have written the
	latter in terms of
	the matrix elements (or $ft$ values)
	of single-$\beta$ decays or charge-exchange reactions,
	calculated in Sec.~\ref{sec:results_beta}.
	
	First, we focus on transitions to the ground state of the final nucleus.
	We can estimate the uncertainty associated with the SSD
	approximation within the ET, 
	as low-lying $1^{+}$ states of the odd-odd system
	are decribed as multiphonon excitations of the lowest $1^{+}$
	state at LO. Thus, their energies and GT matrix elements
	are expected to be
	\begin{align}
		E(1^{+}_{n+1}) & \sim
		E(1^{+}_{1}) + n\omega\,,
		\label{E1n} \\
		M_{\rm GT}\left( 0^{+}_{\rm gs} \rightarrow 1^{+}_{n+1} \right)
		& \sim
		\left(\frac{\omega}{\Lambda}\right)^{n/2}
		M_{\rm GT}\left( 0^{+}_{\rm gs} \rightarrow 1^{+}_{1} \right),
		\label{1nOb0i}
	\end{align}
	according to the power counting introduced in Eq.~(\ref{power}).
	Under these assumptions,
	the uncertainty in the $2\nu\beta\beta$ decay matrix element
	for the transition to the $0^+_{\rm gs}$ state
	due to the SSD approximation scales as
	\begin{equation}
	\begin{split}
	\Delta &M^{2\nu\beta\beta}_{\rm GT} (0^{+}_{\rm gs}\rightarrow 0^{+}_{\rm gs}) \\
	& \sim \sum_{n=1}
	\left(\frac{\omega}{\Lambda}\right)^{n} \frac{
		M_{\rm GT}(1^{+}_{1}\rightarrow 0^{+}_{\rm gs})
		M_{\rm GT}(0^{+}_{\rm gs}\rightarrow 1^{+}_{1})}
	{(D_{10^{+}_{\rm gs}}+n\omega)/m_{e}} \\
	& = \frac{D_{10^{+}_{\rm gs}}}{\Lambda}
	\Phi\left(\frac{\omega}{\Lambda},1,
	\frac{D_{10^{+}_{\rm gs}}+\omega}{\omega}\right)
	M^{2\nu}_{\rm GT}(0^{+}_{\rm gs}\rightarrow 0^{+}_{\rm gs})\,,
	\end{split}
	\label{DeltabbGTme}
	\end{equation}
	where
	\begin{equation}
	\Phi(z,s,a) \equiv \sum_{n=0}^{\infty} \frac{z^{n}}{(a+n)^{s}} \,,
	\end{equation}
	is the Lerch transcendent.
	The relative uncertainty $\delta$ is
	\begin{equation}
	\delta({\rm gs} \rightarrow {\rm gs})= \frac{D_{10^{+}_{\rm gs}}}{\Lambda}
	\Phi\left(\frac{\omega}{\Lambda},1,
	\frac{D_{10^{+}_{\rm gs}}+\omega}{\omega}\right)\,.
	\label{SSD_relative}
	\end{equation}
	Whether this systematic error due to the SSD approximation
	is smaller or larger than the uncertainty associated with the order
	at which the matrix elements are calculated
	depends on the energy scales $\omega$, $\Lambda$ and $D_{10^{+}_{\rm gs}}$.
	
	\begin{figure}[b]
		\centering
		\includegraphics[width=0.8\columnwidth]{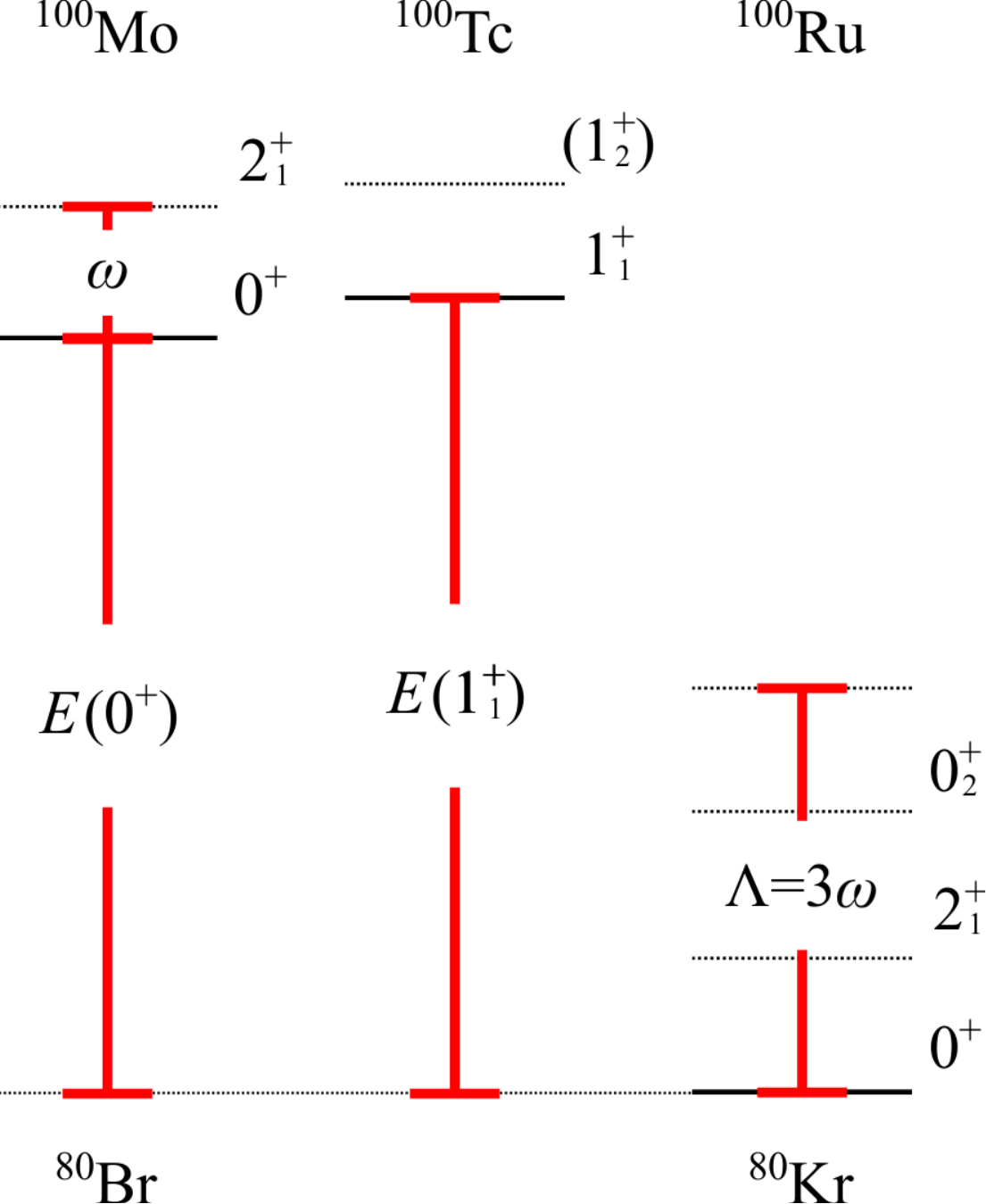}
		\caption{Energy scales relevant to the
			$2\nu\beta\beta$ decay matrix element from the $0^{+}_{\rm gs}$
			ground state of the parent nucleus $^{100}{\rm Mo}$ via the
			$1^+$ intermediate states in $^{100}{\rm Tc}$
			to the $0^{+}_{\rm gs}$ ground state and $0^{+}_{2}$ excited state of the
			daughter nucleus $^{100}{\rm Ru}$.}
		\label{100mo-100ru-bbdecay}
	\end{figure}
	
	Similarly to single-$\beta$ decays,
	we can also calculate
	the matrix elements for $2\nu\beta\beta$ decays
	into $0^{+}_{2}$ excited states within the ET.
	Figure~\ref{100mo-100ru-bbdecay} shows a diagram
	with the relevant energy scales of the nuclei involved.
	In this case, the SSD approximation is not expected to work so well,
	because the contributions from the second and third $1^+$ states ($n=2,3$) in
	\begin{equation}
	M^{2\nu\beta\beta}_{\rm GT}(0^{+}_{\rm gs}\rightarrow 0^{+}_{2})
	= \sum_{n=1} \frac{
		M_{\rm GT}(1^{+}_{n}\rightarrow 0^{+}_{2})
		M_{\rm GT}(0^{+}_{\rm gs}\rightarrow 1^{+}_{n})}
	{D_{n0^{+}_2}/m_{e}}\,,
	\label{01to02}
	\end{equation}
	contain the same number of $d$ operators as the first term.
	Thus, based on the power counting in Eq.~(\ref{power}),
	the first three terms are expected to scale similarly.
	Nevertheless, if only the first term in Eq.~(\ref{01to02}) is considered,
	the $2\nu\beta\beta$ decay matrix element takes the approximate form
	\begin{multline}
		M^{2\nu\beta\beta}_{\rm GT}(0^{+}_{\rm gs} \rightarrow 0^{+}_{2})
		\approx \frac{
			M_{\rm GT}(1^{+}_{\rm gs}\rightarrow 0^{+}_{2})
			M_{\rm GT}(0^{+}_{\rm gs}\rightarrow 1^{+}_{1})}
		{D_{10^{+}_2}/m_{e}} \\
		\approx \frac{D_{10^{+}_{\rm gs}}}{D_{10^{+}_2}}
		\frac{M_{\rm GT}(1^{+}_{1}\rightarrow 0^{+}_{2})}
		{M_{\rm GT}(1^{+}_{1}\rightarrow 0^{+}_{\rm gs})} 
		M^{2\nu}_{\rm GT}(0^{+}_{\rm gs}\rightarrow 0^{+}_{\rm gs})\,,
		\label{SSD_ex}
	\end{multline}
	with a relative uncertainty
	\begin{equation}
	\delta({\rm gs} \rightarrow 0^+_2) = \frac{D_{10^{+}_2}}{D_{20^{+}_2}} + \frac{D_{10^{+}_2}}{D_{30^{+}_2}}
	+ \frac{D_{10^{+}_2}}{\Lambda} \Phi\left(\frac{\omega}{\Lambda},1,
	\frac{D_{30^{+}_2}+\omega}{\omega} \right).
	\label{deltaSSD_ex}
	\end{equation}

	\begin{figure*}[]
		\centering
		\includegraphics[width=0.65\textwidth]{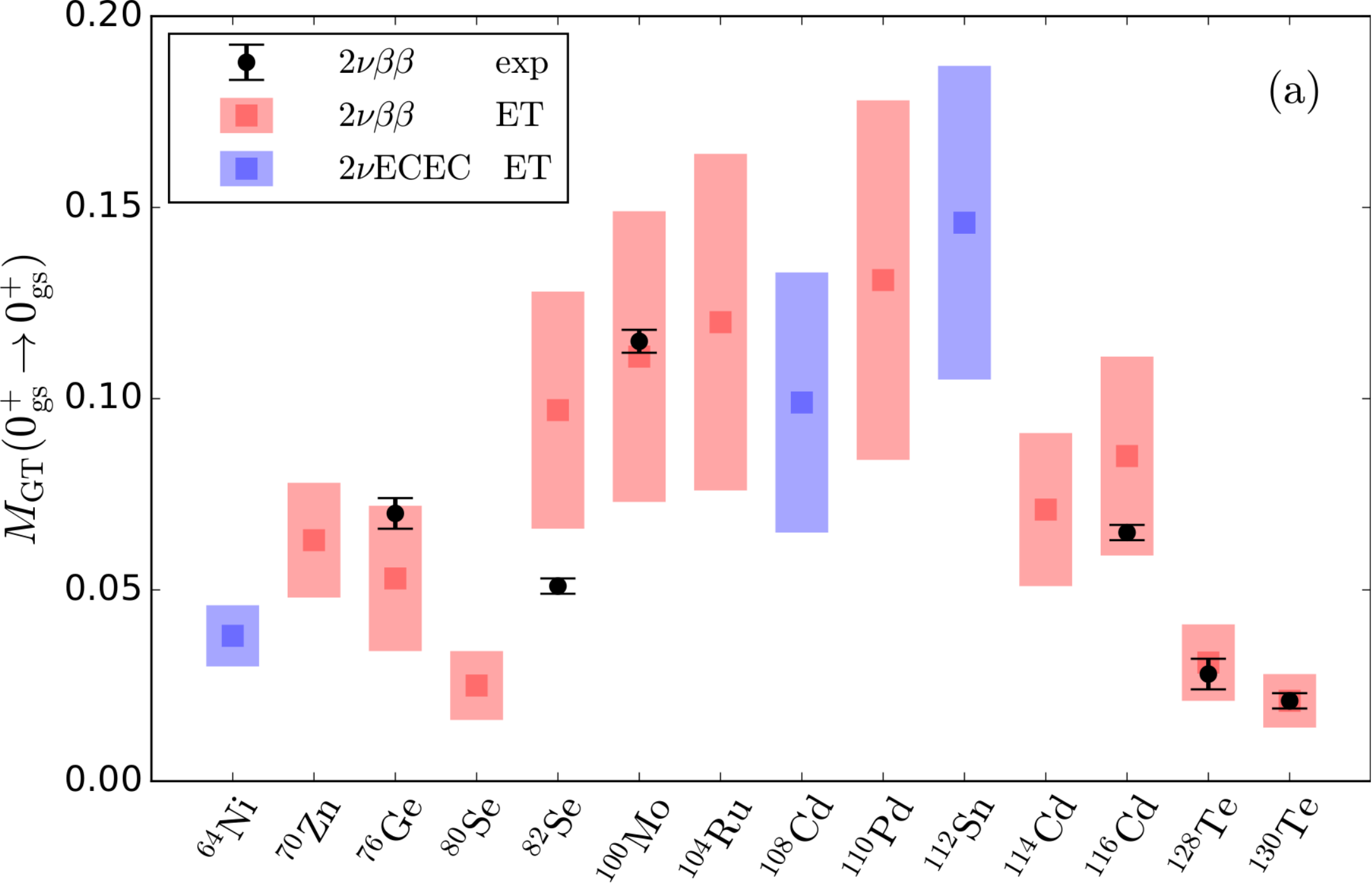}
		\includegraphics[width=0.65\textwidth]{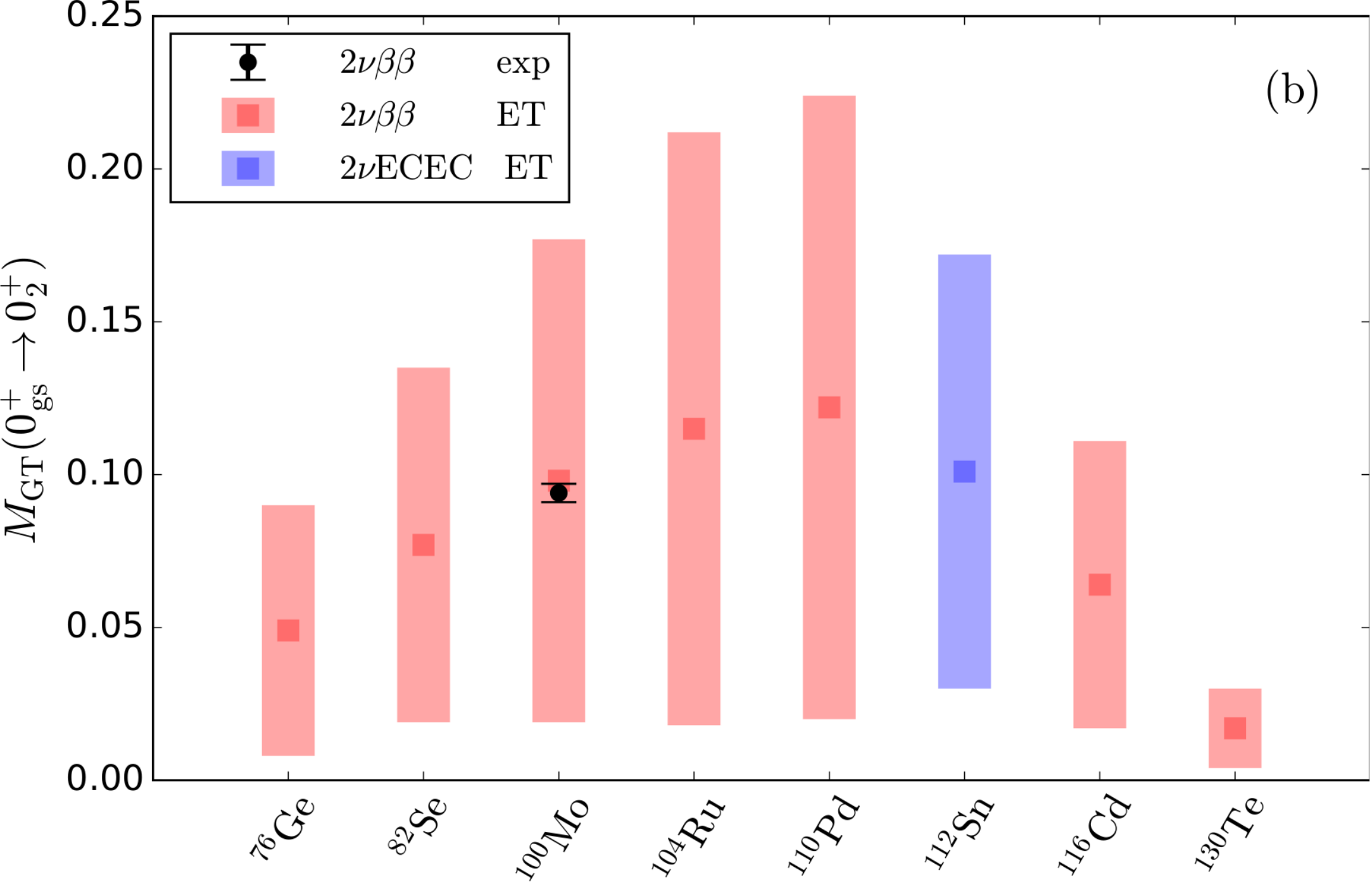}
		\includegraphics[width=0.65\textwidth]{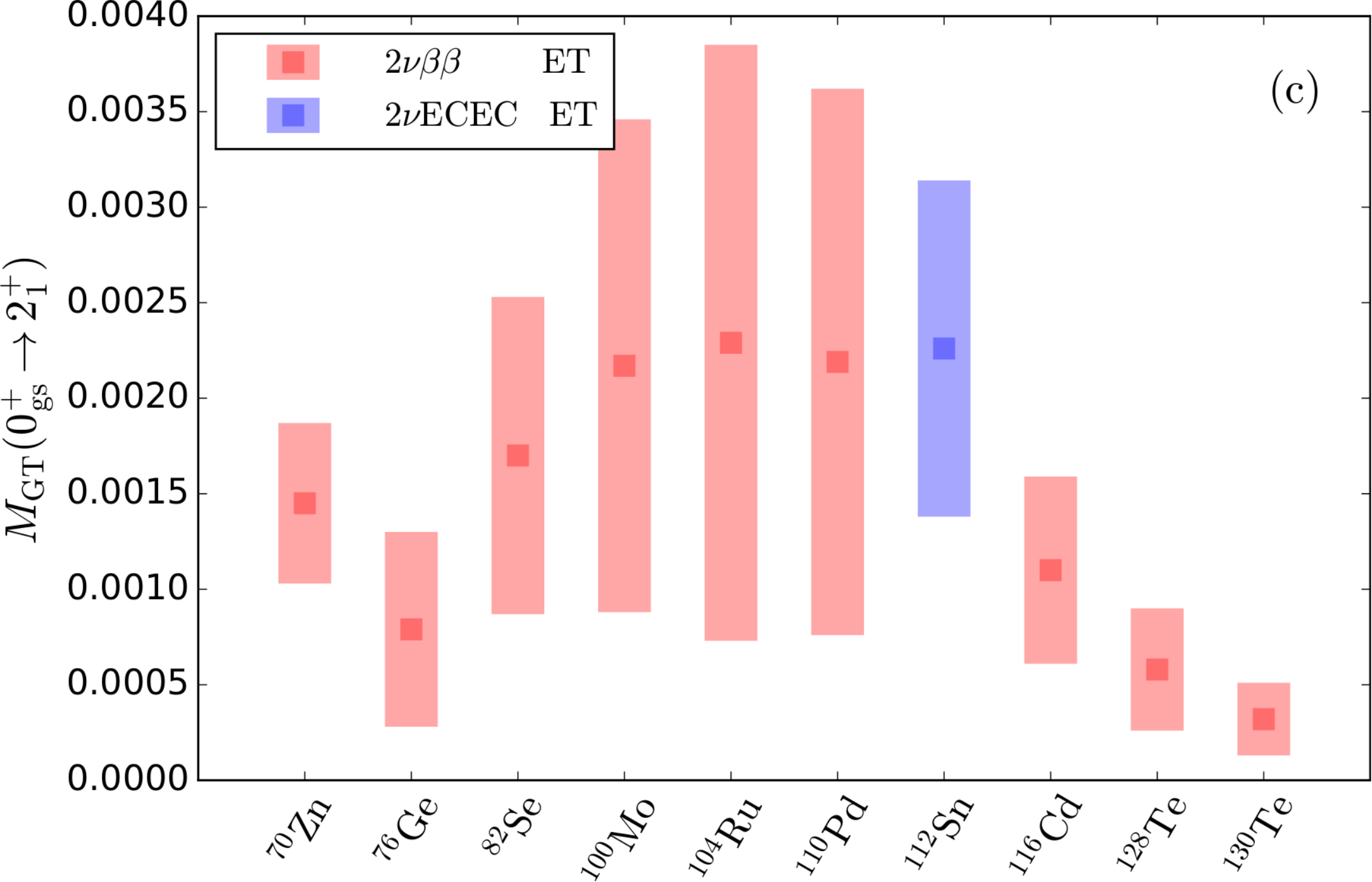}
		\caption{Calculated ET matrix elements for $2\nu\beta\beta$ (red bands)
			and $2\nu$ECEC (blue bands) decays to low-lying collective ground (a) and excited $0^+_2$ (b) and $2^+_1$ (c) states of the daughter
			nuclei, in comparison with experiment (black bars).
			The LECs of the ET are fitted to
			single-$\beta$ and/or EC decays (or to GT strengths if the former are not available)
			The experimental matrix elements are taken from Ref.~\cite{barabash2015}.}
		\label{GT2nu}
	\end{figure*}
	
	We can reduce this relative uncertainty
	assuming that the contributions due to the first three terms are in phase.
	This yields the following matrix element
	\begin{equation}
	\begin{split}
	M^{2\nu\beta\beta}_{\rm GT}(0^{+}_{\rm gs} & \rightarrow 0^{+}_{2}) \approx 
	\left(1 + \frac{D_{10^{+}_2}}{D_{20^{+}_2}} + \frac{D_{10^{+}_2}}{D_{30^{+}_2}} \right) \\
	\times \frac{D_{10^{+}_{\rm gs}}}{D_{10^{+}_2}} &
	\frac{M_{\rm GT}(1^{+}_{1}\rightarrow 0^{+}_{2})}
	{M_{\rm GT}(1^{+}_{1}\rightarrow 0^{+}_{\rm gs})} 
	M^{2\nu\beta\beta}_{\rm GT}(0^{+}_{\rm gs}\rightarrow 0^{+}_{\rm gs})\,,
	\end{split}
	\label{TSA}
	\end{equation}
	and the reduced relative uncertainty
	\begin{equation}
	\begin{split}
	\delta({\rm gs} \rightarrow 0^+_2) = & \frac{\omega}{\Lambda}
	\left(\frac{D_{10^{+}_2}}{D_{20^{+}_2}} + \frac{D_{10^{+}_2}}{D_{30^{+}_2}}\right) \\
	& + \frac{D_{10^{+}_2}}{\Lambda} \Phi\left(\frac{\omega}{\Lambda},1,
	\frac{D_{30^{+}_2}+\omega}{\omega} \right) .
	\end{split}
	\label{deltaTSA}
	\end{equation}
	In Sec.~\ref{sec:results_2nbb}, we compare to experimental results the
	$2\nu\beta\beta$ matrix element of $^{100}{\rm Mo}$ decaying into the $0^+_2$ state of
	$^{100}{\rm Ru}$ using Eqs.~(\ref{SSD_ex}) and (\ref{TSA}), with the uncertainties given by Eqs.~(\ref{deltaSSD_ex}) and (\ref{deltaTSA}).
	
	The ET can also predict $2\nu\beta\beta$ decays matrix elements
	to excited $2^{+}_1$ states of the daughter nucleus.
	Here, because energy denominators appear to the third
	power, we only consider the contribution due to the
	first intermediate $1^+$ state. Then, matrix elements for decays
	into $2^+_1$ states take the approximate form
	\begin{equation}
	\begin{split}
	M^{2\nu\beta\beta}_{\rm GT} &
	(0^{+}_{\rm gs}\rightarrow 2^{+}_1) \\
	\approx & \sqrt{\frac{1}{3}}
	\frac{m_e^2 D_{10^{+}_{\rm gs}}}{D_{12^{+}_{1}}^3}
	\frac{M_{\rm GT}(1^{+}_{1}\rightarrow 2^{+}_1)}
	{M_{\rm GT}(1^{+}_{1}\rightarrow 0^{+}_{\rm gs})}
	M^{2\nu\beta\beta}_{\rm GT}
	(0^{+}_{\rm gs}\rightarrow 0^{+}_{\rm gs}),
	\end{split}
	\label{TSA2}
	\end{equation}
	with an associated relative uncertainty given by
	\begin{equation}
	\begin{split}
	\delta({\rm gs} \rightarrow 2^+_1) = & \frac{D_{12^{+}_1}^3}{\omega^3} \Phi\left(\frac{\omega}{\Lambda},3,
	\frac{D_{12^{+}_1}+\omega}{\omega} \right) .
	\end{split}
	\label{deltaTSA2}
	\end{equation}
	This uncertainty
	varies from nucleus to nucleus,
	depending on the energy scales $\omega$ and $D_{12^+_1}$.
	
	\subsection{Results for \texorpdfstring{$\boldsymbol{2\nu\beta\beta}$}{} decay
		\label{sec:results_2nbb}}
	
	As an example, let us first focus on
	the $^{100}{\rm Mo}$ $2\nu\beta\beta$ decay.
	The relevant energy scales are shown in Fig.~\ref{100mo-100ru-bbdecay}.
	Setting the energy scale $\omega$ equal to the average of the excitation energies
	of the $2^{+}_{1}$ states in the even-even nuclei
	and the breakdown scale to $\Lambda=3\omega$,
	Eqs.~(\ref{SSD}) and~(\ref{SSD_relative})
	yield for the decay to the $^{100}{\rm Ru}$ ground state 
	\begin{equation}
	M^{2\nu\beta\beta}_{\rm GT}(0^{+}_{\rm gs}\rightarrow 0^{+}_{\rm gs})
	\approx 0.111(38) \,,
	\end{equation}
	where we have fitted the LECs of the ET to the experimental single-$\beta$ and EC decays
	and the uncertainty given is thus dominated by the SSD approximation from Eq.~(\ref{SSD_relative}).
	We note that the uncertainty due to the SSD approximation (35\% in this case)
	is of the same order as the uncertainty associated with the effective
	nuclear states and GT operator used to calculate
	the single-$\beta$ decay matrix elements at LO.
	Therefore, the SSD approximation is appropriate to obtain
	$2\nu\beta\beta$ decay matrix elements at LO.
	
	\begin{table*}[t]
		\caption{$2\nu\beta\beta$ and $2\nu$ECEC matrix elements
			for the decays to low-lying collective states of the daughter nuclei.
			The values for the LEC $\omega$ are listed in the second column. The LECs of the effective GT operator are fitted to reproduce the
			matrix elements for single-$\beta$ and/or EC decays, or GT strengths from charge-exchange reactions if the former are not available
			(third and fourth columns). Calculated decays into ground states (sixth column) assume the SSD approximation,
			while decays to $0^+_2$ (eighth column) and $2^+_1$  (nineth column) excited states are calculated according to Eqs.~(\ref{TSA}) and~(\ref{deltaTSA}) and Eqs.~(\ref{TSA2}) and~(\ref{deltaTSA2}), respectively. 
			Experimental data for decays to ground (fifth column) and excited (seventh column) $0^+$ states are taken from Ref.~\cite{barabash2015}.
			}
		\centering
		\begin{tabular}{c | c | c c | c c | c c | c }
			\hline\hline
			${\rm Decay}$ & $\omega [{\rm keV}]$ &
			\multicolumn{2}{c |}{$M_{\rm GT}$} &
			\multicolumn{2}{c |}{$M^{2\nu\beta\beta/{\rm ECEC}}_{\rm GT}(0^+_{\rm gs}\rightarrow 0^+_{\rm gs})$} &
			\multicolumn{2}{c |}{$M^{2\nu\beta\beta/{\rm ECEC}}_{\rm GT}(0^+_{\rm gs}\rightarrow 0^+_2)$} &
			$M^{2\nu\beta\beta/{\rm ECEC}}_{\rm GT}(0^+_{\rm gs}\rightarrow 2^+_1)$ \\
			& &
			$(0^{+}_{\rm gs}\!\!\rightarrow\!1^{+}_{1})$ &
			$(1^{+}_{1}\!\!\rightarrow\!0^{+}_{\rm gs})$ &
			Expt. & ET & Expt. & ET & ET \\
			\hline
			$^{64}{\rm Zn}\overset{\rm ECEC}
			{\longrightarrow}{}^{64}{\rm Ni}$ & 
			$1168.7$ & $0.239$ & $0.350$ &
			& $0.038(8)$ &
			& &
			\\
			$^{70}{\rm Zn}\overset{\beta^{-}\beta^{-}}
			{\longrightarrow}{}^{70}{\rm Ge}$ & 
			$962.2$ & $0.467$ & $0.304$ &
			& $0.063(15)$ &
			& &
			$0.00145(42)$ \\
			$^{76}{\rm Ge}\overset{\beta^{-}\beta^{-}}
			{\longrightarrow}{}^{76}{\rm Se}$ & 
			$561.0$ & $0.456$\footnotemark[1] & $0.456$\footnotemark[2] &
			$0.070(4)$ & $0.053(19)$ &
			& $0.049(41)$ &
			$0.00079(51)$ \\
			$^{80}{\rm Se}\overset{\beta^{-}\beta^{-}}
			{\longrightarrow}{}^{80}{\rm Kr}$ & 
			$641.4$ & $0.494$ & $0.193$ &
			& $0.025(9)$ &
			& \\
			$^{82}{\rm Se}\overset{\beta^{-}\beta^{-}}
			{\longrightarrow}{}^{82}{\rm Kr}$ & 
			$715.6$ & $0.581$\footnotemark[1]  & $0.548$\footnotemark[3] &
			$0.051(2)$ & $0.097(31)$ &
			& $0.077(58)$ &
			$0.00170(83)$ \\
			\hline
			$^{100}{\rm Mo}\overset{\beta^{-}\beta^{-}}
			{\longrightarrow}{}^{100}{\rm Ru}$ &
			$537.5$ & $0.675$ & $0.542$ &
			$0.115(3)$ & $0.111(38)$ &
			$0.094(3)$ & $0.098(79)$ &
			$0.00217(129)$ \\
			$^{104}{\rm Ru}\overset{\beta^{-}\beta^{-}}
			{\longrightarrow}{}^{104}{\rm Pd}$ &
			$456.9$ & $0.740$ & $0.568$ &
			& $0.120(44)$ &
			& $0.115(97)$ &
			$0.00229(156)$ \\
			$^{106}{\rm Cd}\overset{\rm ECEC}
			{\longrightarrow}{}^{106}{\rm Pd}$ &
			$572.2$ & $<\!0.953$ & $0.371$ &
			& $<\!0.114(38)$ &
			& $<\!0.097(76)$ &
			$<\!0.00243(133)$ \\
			$^{108}{\rm Cd}\overset{\rm ECEC}
			{\longrightarrow}{}^{108}{\rm Pd}$ &
			$533.4$ & $0.655$ & $0.478$ &
			& $0.099(34)$ &
			& &
			\\
			$^{110}{\rm Pd}\overset{\beta^{-}\beta^{-}}
			{\longrightarrow}{}^{110}{\rm Cd}$ &
			$515.8$ & $0.964$ & $0.500$ &
			& $0.131(47)$ &
			& $0.122(102)$ &
			$0.00219(143)$ \\
			$^{112}{\rm Sn}\overset{\rm ECEC}
			{\longrightarrow}{}^{112}{\rm Cd}$ &
			$937.1$ & $0.910$ & $0.512$ &
			& $0.146(41)$ &
			& $0.101(71)$ &
			$0.00226(88)$ \\
			$^{114}{\rm Cd}\overset{\beta^{-}\beta^{-}}
			{\longrightarrow}{}^{114}{\rm Sn}$ &
			$929.2$ & $0.384$ & $0.622$ &
			& $0.071(20)$ &
			& &
			\\
			$^{116}{\rm Cd}\overset{\beta^{-}\beta^{-}}
			{\longrightarrow}{}^{116}{\rm Sn}$ &
			$903.5$ & $0.622$ & $0.499$ &
			$0.065(2)$ & $0.085(26)$ &
			& $0.064(47)$ &
			$0.00110(49)$ \\
			\hline
			$^{128}{\rm Te}\overset{\beta^{-}\beta^{-}}
			{\longrightarrow}{}^{128}{\rm Xe}$ &
			$593.0$ & $0.319$ & $0.319$\footnotemark[4] &
			$0.028(4)$ & $0.031(10)$ &
			& &
			$0.00058(32)$ \\
			$^{130}{\rm Te}\overset{\beta^{-}\beta^{-}}
			{\longrightarrow}{}^{130}{\rm Xe}$ &
			$687.8$ & $0.269$\footnotemark[1] & $0.269$\footnotemark[5] &
			$0.021(2)$ & $0.021(7)$ &
			& $0.017(13)$ &
			$0.00036(19)$ \\
			\hline\hline
		\end{tabular}
		\footnotetext[1]{Calculated with LECs fitted to charge-exchange reactions.}
		\footnotetext[2]{Assumed similar to the
			$^{76}{\rm Ge}\rightarrow{}^{76}{\rm As}$ matrix element.}
		\footnotetext[3]{Assumed similar to the
			$^{82}{\rm Rb}\rightarrow{}^{82}{\rm Kr}$ matrix element.}
		\footnotetext[4]{Assumed similar to the
			$^{128}{\rm I}\rightarrow{}^{128}{\rm Te}$ matrix element.}
		\footnotetext[5]{Assumed similar to the
			$^{130}{\rm I}\rightarrow{}^{130}{\rm Te}$ matrix element.}
		\label{bbGTme}
	\end{table*}
	
	Figure~\ref{GT2nu} shows the ET results for the $2\nu\beta\beta$ and
	$2\nu$ECEC decays of several nuclei with mass number from $A=64$ to 
	$A=130$. The LECs of the ET are again fitted to experimental single-$\beta$ 
	and/or EC decays, or to GT strengths if the former are not available.
	The results, as well as the GT matrix elements used for both 
	single-$\beta$ decay branches, are also given in Table~\ref{bbGTme}.
	In some cases, for which there is no experimental data
	on single-$\beta$ decay, EC decay or GT strengths,
	the GT matrix elements were assumed to be similar for both
	$\beta$ decay branches, as indicated in Table~\ref{bbGTme}.
	The similarity of the two matrix elements is a prediction of the ET.
	The top panel in Fig.~\ref{GT2nu} shows that the theoretical results with uncertainties
	 for decays to the ground state of the daughter nucleus agree remarkably well with experimental data when available.
	We provide additional ET predictions
	for the kinematically less favored $2\nu\beta\beta$ decays and $2\nu$ECEC transitions.
	While the $2\nu\beta\beta$ decays from $^{48}{\rm Ca}$,
	$^{96}{\rm Zr}$, $^{136}{\rm Xe}$, and $^{150}{\rm Nd}$ have been measured,
	they are not included in our comparison because their low-energy properties,
	or those of the corresponding daughter nuclei,
	do not resemble those expected for collective spherical systems.
	
	The middle panel of Fig.~\ref{GT2nu} shows predicted ET matrix elements
	for $2\nu\beta\beta$ decays into energetically allowed $0^+_2$ states. The matrix elements, calculated with Eq.~(\ref{TSA}), are listed in Table~\ref{bbGTme}. The ET uncertainties are larger than those for decays to the ground state because of the larger impact of high-energy $1^+$ states. Nonetheless, the predicted matrix elements for transitions to excited $0^+_2$ and the ground state are similar.
	
	It is especially interesting to compare to the measured $^{100}{\rm Mo}$
	decay into the $0^{+}_{2}$ state of $^{100}{\rm Ru}$.
	In this case the SSD approximation given by Eqs.~(\ref{SSD_ex})
	and~(\ref{deltaSSD_ex}) yields
	\begin{equation}
	M^{2\nu\beta\beta}_{\rm GT}(0^{+}_{\rm gs}\rightarrow 0^{+}_{2}) \approx  0.040(71)\,,
	\label{ru_ex1}
	\end{equation}
	which clearly shows (the relative error reaches 180\%)
	that the contributions from the second and third $1^+$ states
	in the sum in Eq.~(\ref{01to02}) cannot be neglected.
	When we take these two additional terms into account
	and assume that the first three contributions add up with the same phase,
	the $2\nu\beta\beta$ matrix element given by Eqs.~(\ref{TSA})
	and~(\ref{deltaTSA}) becomes
	\begin{equation}
	M^{2\nu\beta\beta}_{\rm GT}(0^{+}_{\rm gs}\rightarrow 0^{+}_{2})\approx 0.098(79)\,,
	\label{ru_ex2}
	\end{equation}
	with a smaller but still significant relative uncertainty of 80\%.
	We note that the two ET results
	are consistent with the experimental value
	$M^{2\nu\beta\beta}_{\rm GT}(0^+_{\rm gs}\rightarrow
	0^+_2)=0.094(3)$. The agreement with the result yielded by
	Eq.~(\ref{TSA}) suggests that for this
	transition the contributions from the first three $1^+$ states
	in Eq.~(\ref{01to02}) likely add up in phase.

	The bottom panel of Fig.~\ref{GT2nu} shows the ET predictions
	for matrix elements of decays into excited
	$2^+_1$ states, calculated using Eqs.~(\ref{TSA2}) and (\ref{deltaTSA2}). 
	The values with corresponding uncertainties are listed in Table~\ref{bbGTme}.
	The higher power of the energy denominator compared to decays to $0^+$ states suppresses the matrix elements into $2^+_1$ states. Not a single transition of this type has been detected experimentally. On the other hand, the expected dominance of the transition through the lowest $1^+$ state in the intermediate nucleus reduces the ET uncertainties with respect to transitions to excited $0^+_2$ states.

	The LO $2\nu\beta\beta$ and $2\nu$ECEC results 
	can be systematically improved, and the uncertainties reduced, at higher orders in the ET. The cost is to introduce additional LECs that need to be fitted to data on the energy spectra, $\beta^-$ and EC decays of the intermediate odd-odd nuclei.
	
	\section{Summary and outlook}
	\label{summary}
	
	We have studied $\beta$ and EC decays
	within an ET that treats nuclei as a spherical even-even core
	that can be coupled to one additional neutron and/or proton.
	By fitting the LEC of the effective GT operator to the decay
	to the ground state of the daughter nucleus,
	the matrix elements corresponding to the decays to collective excited states
	of the same nucleus are predicted by the ET.
	We have also used experimental data on charge-exchange reactions
	to fit the LECs and predict the decay into the daughter ground state.
	One of the advantages of the ET is that it provides a power counting
	that allow us to estimate the theoretical uncertainty of the calculations.
	When this uncertainty is included, we find good agreement
	between the ET predictions and experiment.
	The consistency covers more than 20 spherical medium-mass and heavy nuclei.
	Our results thus suggest that transitions due to the weak interactions
	can be well described by the ET at LO.
	
	In addition, we have used data on single-$\beta$ and/or EC decays, or
	GT strengths from charge-exchange reactions,
	to calculate matrix elements for $2\nu\beta\beta$ and $2\nu$ECEC decays.
	We generally assume the SSD approximation,
	which is consistent with obtaining results with the ET at LO.
	Without modifying the LECs of the first-order processes,
	the ET gives $2\nu\beta\beta$ decay matrix elements
	consistent with experiment. In one case the agreement extends
	to the $2\nu\beta\beta$ decay into an excited $0^+_2$ state. 
	Furthermore, we have predicted several 
	$2\nu\beta\beta$ and $2\nu$ECEC matrix elements into ground and excited $0^+_2$ and $2^+_1$ states.	
	Based on the power counting of the ET, the LO matrix elements to ground and excited states are predicted consistently.
	The validity of the ET will be challenged when future experiments measure these transitions.
	
	Future work includes a more precise ET calculation. This will require us to
	include higher-order corrections to the Hamiltonian for
	the odd-odd nuclei, which will in turn correct our LO approximation for
	the low-lying odd-odd states, as well as higher-order corrections to the effective GT operator.
	Based on the power counting, we expect these two kinds of corrections
	to contribute a factor of $\omega/\Lambda \sim 30\%$ to the reduced GT matrix elements for $2\nu\beta\beta$ decays.
	Work in these directions is in progress.
	
	In addition, it will be important
	to perform LO calculations within ETs
	for axial~\cite{papenbrock2011, zhang2013, papenbrock2014,
	coelloperez2015-1} and triaxial~\cite{chen2017} nuclei. This
	will provide access to nuclei whose ground states deviate
	from sphericity, and is thus also expected to improve the
	calculations for some nuclei in this work, whose ground
	states possess some deformation.
	
	An extension of the ET presented here
	is also a promising framework to estimate the matrix elements
	of $0\nu\beta\beta$ decays with theoretical uncertainties.
	For this purpose, the effective $0\nu\beta\beta$ decay operator
	needs to be written in terms of the DOF of the ET,
	and the corresponding LECs would have to be fixed.
	Since there is no experimental data on $0\nu\beta\beta$ decay yet,
	the fitting of the LECs can be validated against
	other experimental data strongly correlated to $0\nu\beta\beta$ decay.
	Alternatively, the ET LECs can be fitted to existing nuclear structure calculations,
	which is similar to the strategy followed by interacting boson model 
	calculations (see, e.g., Ref.~\cite{barea2009}).
	Both strategies will provide
	constraints and predictions for the $0\nu\beta\beta$ matrix elements
	with ET uncertainties.
	
	\begin{acknowledgments}
		We thank T.\ Papenbrock for very useful comments on the manuscript.
		This work was supported in part by
		the Deutsche Forschungsgesellschaft under Grant SFB 1245,
		MEXT, as Priority Issue on Post-K Computer
		(Elucidation of the Fundamental Laws and Evolution of the Universe), and JICFuS.
	\end{acknowledgments}

\end{document}